\documentclass[prx,aps,twocolumn,eqsecnum,superscriptaddress]{revtex4-1}
\usepackage{amssymb}
\usepackage{amsmath}
\usepackage{graphicx}
\usepackage[colorlinks=true,linkcolor=blue,anchorcolor=red,citecolor=blue,urlcolor=blue]{hyperref}
\usepackage{amsmath}

\begin{document}
%\begin{CJK}{GBK}{kai}

\title{Quantum Transport in Topological Semimetals under Magnetic Fields}

\author{Hai-Zhou Lu}
\affiliation{Department of Physics, South University of Science and Technology of China, Shenzhen 518055, China}

\author{Shun-Qing Shen}
\affiliation{Department of Physics, The University of Hong Kong, Pokfulam Road, Hong Kong, China}

\date{\today }

\begin{abstract}
Topological semimetals are three-dimensional topological states of matter,
in which the conduction and valence bands touch at a finite number of points, i.e., the Weyl nodes. Topological semimetals host paired monopoles and antimonopoles of Berry
curvature at the Weyl nodes and topologically protected Fermi arcs at certain surfaces. We review our recent works on quantum transport in topological semimetals, according to the strength of the magnetic field. At weak magnetic fields, there are competitions between the positive magnetoresistivity induced by the weak anti-localization effect and negative magnetoresistivity related to the nontrivial Berry curvature. We propose a fitting formula for the magnetoconductivity of the weak anti-localization. We expect that the weak localization may be induced by inter-valley effects and interaction effect, and occur in double-Weyl semimetals. For the negative magnetoresistance induced by the nontrivial Berry curvature in topological semimetals, we show the dependence of the negative magnetoresistance on the carrier density. At strong magnetic fields, specifically, in the quantum limit, the magnetoconductivity depends on the type and range of the scattering potential of disorder.
The high-field positive magnetoconductivity may not be a compelling signature of the chiral anomaly. For long-range Gaussian scattering potential and half filling, the magnetoconductivity can be linear in the quantum limit. A minimal conductivity is found at the Weyl nodes although the density of states vanishes there.
\end{abstract}

\maketitle



\tableofcontents

\section{Introduction}

Weyl semimetal is a three-dimensional (3D) topological
state of matter, in which the conduction and valence energy bands
touch at a finite number of nodes \cite{Balents11physics,Volovik03book}. The nodes
always appear in pairs, in each pair the quasiparticles carry opposite chirality and linear dispersion, much like
a 3D analog of graphene. In the past few
years, a number of materials have been suggested to
host Weyl fermions \cite{Wan11prb,Yang11prb,Burkov11prl,Xu11prl,Delplace12epl,Jiang12pra,Young12prl,Wang12prb,Singh12prb,Wang13prb,LiuJP14prb,Bulmash14prb}.
The topological semimetals can be simply classified into Weyl semimetals and Dirac semimetals. In a Weyl semimetal, each Weyl node is non-degenerate, while in a Dirac semimetal, the Weyl nodes are degenerate due to time-reversal and inversion symmetry \cite{Young12prl}. Recently, angle-resolved photoemission spectroscopy
(ARPES) has identified the Dirac nodes in (Bi$_{1-x}$In$_{x}$)$_{2}$Se$_{3}$ \cite{Brahlek12prl,Wu13natphys},
Na$_{3}$Bi \cite{Wang12prb,Liu14sci,Wang13prb,Xu15sci}, Cd$_{3}$As$_{2}$
\cite{Wang13prb,Liu14natmat,Neupane14nc,Yi14srep,Borisenko14prl},
and Weyl nodes in the TaAs family \cite{Weng15prx,Huang15nc,Lv15prx,Xu15sci-TaAs} and YbMnBi$_2$ \cite{Borisenko15arXiv}.

The monopoles hosted by topological semimetals may
lead to a number of novel transport effects \cite{Nielsen81npb,Nielsen83plb,Son13prb,Burkov14prl-chiral,Kharzeev13prb,Parameswaran14prx,ZhouJH15prb,Wan11prb,Yang11prb,Burkov11prl,Xu11prl,Son12prl,Stephanov12prl,Landsteiner11prl,Chang15prb,JiangQD15prl,JiangQD16prb,ChenCZ15prl,ChenCZ16prb}, including the ``chiral anomaly" \cite{Nielsen81npb,Nielsen83plb,Son13prb,Burkov14prl-chiral,Kharzeev13prb,Parameswaran14prx,ZhouJH15prb}, the anomalous Hall effect \cite{Wan11prb,Yang11prb,Burkov11prl,Xu11prl}, the chiral magnetic effect \cite{Son12prl,Stephanov12prl,Landsteiner11prl,Chang15prb}. There have been growing efforts exploring the transport in topological semimetals, including
Bi$_{0.97}$Sb$_{0.03}$ \cite{Kim13prl,Kim14prb}, ZrTe$_5$ \cite{Li16np,ChenRY15prl,ZhengGL16prb}, Na$_3$Bi \cite{Xiong15sci}, Cd$_3$As$_2$ \cite{Jeon14nmat,Liang15nmat,Feng15prbrc,He14prl,Zhao15prx,Cao15nc,Shekhar15np,Narayanan15prl,LiCZ15nc,LiH16nc,ZhangC15arXiv,WangH16nmat,Aggarwal16nmat}, TaAs \cite{HuangXC15prx,ZhangCL16nc}, TaP \cite{ZhangCL15prb-TaP,Arnold16nc,ZhangCL15arXiv-TaP}, NbAs \cite{YangXJ15arXiv,YangXJ15arXiv-NbAs}, NbP \cite{Shekhar15np,WangZ16prb}, HfTe$_{5}$ \cite{WangHC16prb}, etc.

Study of magneto-transport properties is one of the research focuses in Weyl semimetals.
According to the strength of the magnetic field, the transport in topological semimetals can be classified into four regimes. (i) Near zero field, one has a positive magnetoresistance from the weak anti-localization effect. (ii) At weak parallel magnetic fields, there is a negative magnetoresistance arising from the nontrivial Berry curvature in topological semimetals. (iii) At intermediate magnetic fields, there is the quantum oscillation of resistivity due to the Landau quantization of energy states. (iv) At strong magnetic field, specifically, when only the lowest Landau band is occupied, it is controversial whether a negative magnetoresistance can be regarded as a signature for the chiral anomaly. Also, in most experiments, there is large magnetoresistance in perpendicular magnetic fields, sometimes linearly increases with the field.

In this paper, we review our recent efforts on the quantum transport in topological semimetals \cite{Lu15Weyl-Localization,Lu15Weyl-shortrange,LiH16nc,Dai16prbrc,ZhangSB16njp,WangCM16arXiv}. Part of the contents has been reviewed in Refs. \cite{Lu14spie,Lu16cpb}, where the focus was the weak localization and anti-localization effects. There have been several review articles on topological semimetals \cite{Hosur13Physique}. In Sec. \ref{Sec:Model}, we introduce the models we used for topological semimetals. In Sec. \ref{Sec:WAL}, we summarize the theories of the weak anti-localization for Weyl semimetals and weak localization for double-Weyl semimetals. We propose a formula for the magnetoconductivity induced by the weak (anti-)localization, which is not only applicable for topological semimetals but also for other 3D systems. We also show the weak localization of Weyl fermions as a result of electron-electron interactions and inter-valley effects. In Sec. \ref{Sec:NMR}, we review the experiments on the negative magnetoresistance in topological semimetals, and show the relation between the magnetic monopole and the negative magnetoresistance. In Sec. \ref{Sec:QL}, we review our results on the magnetoconductivity in the quantum limit. Finally, remarks and perspective are given in Sec. \ref{Sec:remark}.

\begin{figure}[htbp]
\centering \includegraphics[width=0.48\textwidth]{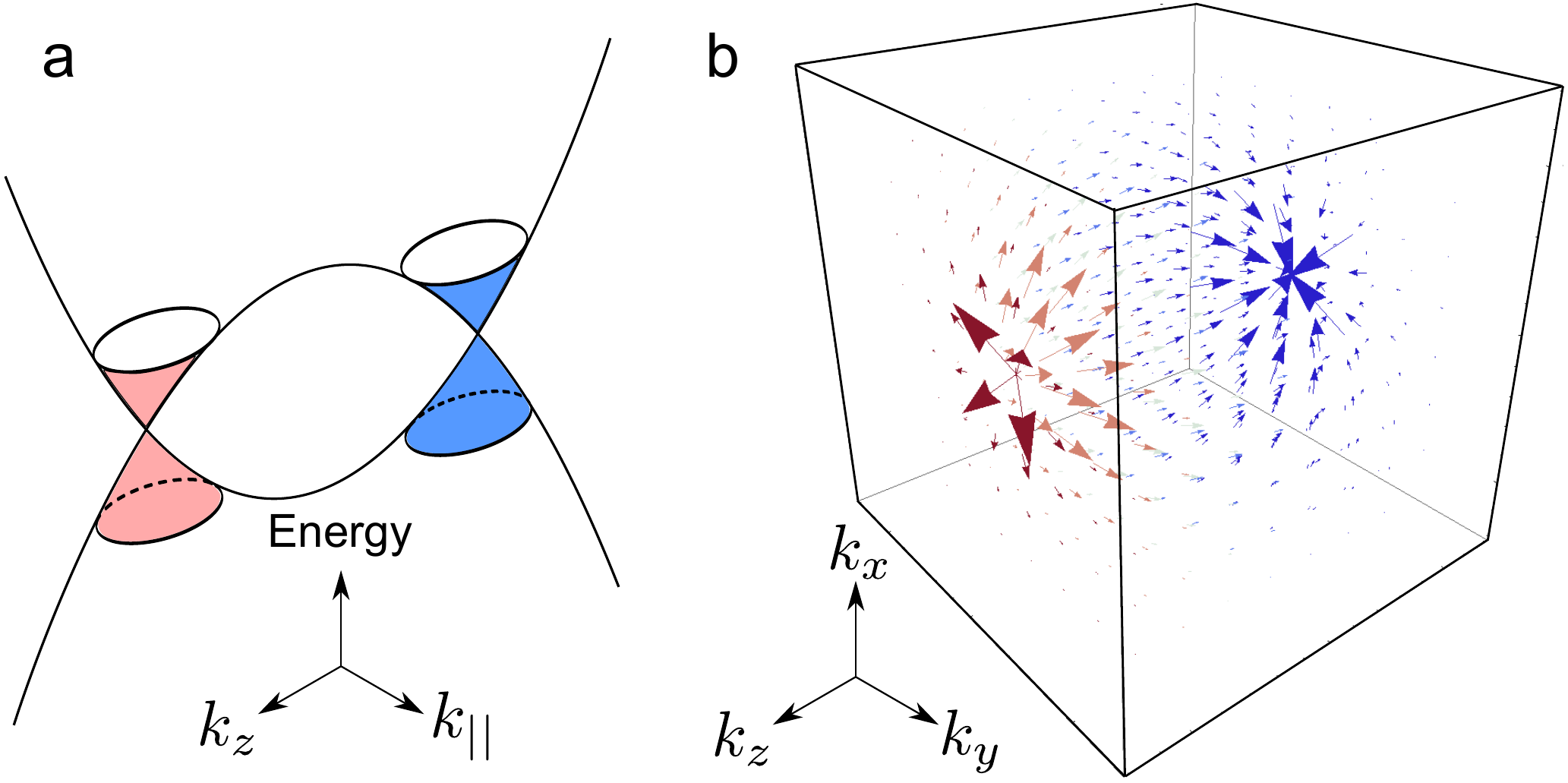} \caption{Nontrivial band structure and Berry curvature of a topological semimetal. (a) A schematic of the energy spectrum of a topological semimetal. $(k_x, k_y, k_z)$ is the wave vector. $k_{||}^2=k_x^2+k_y^2$. (b) The vector plot of the Berry curvature in momentum space. The conduction and valence bands of a topological semimetal touch at the Weyl nodes, at which a pair of monopoles are hosted. The arrows show that the flux of the Berry curvature flows from one monopole (red) to the other (blue), defining the nontrivial topological properties of a topological semimetal. Adapted from Ref. \cite{LiH16nc}.}
\label{Fig:band}
\end{figure}

\section{Effective models}\label{Sec:Model}

\subsection{Two-node model of Weyl semimetal}\label{Sec:two-node-model}

A minimal model for a Weyl semimetal can be written as
\begin{eqnarray}\label{eq:model}
H=A(k_{x}\sigma_{x}+k_{y}\sigma_{y})+\mathcal{M_{\mathbf{k}}}\sigma_{z},
\end{eqnarray}
where $\sigma$ are the Pauli matrices, $\mathcal{M}_{\mathbf{k}}=M_{0}-M_{1}(k_{x}^{2}+k_{y}^{2}+k_{z}^{2})$,
$\mathbf{k}=(k_{x},k_{y},k_{z})$ is the wave vector, and $A$, $M_{0/1}$ are
model parameters. This minimal model gives a global description of
a pair of Weyl nodes of opposite chirality and all the topological
properties. It has an identical structure as that for A-phase of $^{3}$He
superfluids \cite{Shen12book}. If $M_{0}M_{1}>0$, the two bands intersect at $(0,0,\pm k_w)$
with $k_w\equiv\sqrt{M_{0}/M_{1}}$ (see Fig.~\ref{Fig:band}), giving rise to the topological semimetal phase. In the topological semimetal phase, the model can also be written as
\begin{eqnarray}\label{Eq:model}
H =A(k_x\sigma_x+k_y\sigma_y)+M (  k_w^2-\mathbf{k}^2)\sigma_z,
\end{eqnarray}
where $A$, $M$, $k_w$ are
model parameters. The dispersions of two energy bands of
this model are
\begin{eqnarray}
E_{\pm}=\pm\sqrt{[M(k_w^2-\mathbf{k}^2)]^{2}+A^{2}(k_{x}^{2}+k_{y}^{2})},
\end{eqnarray}
which reduce to $E_{\pm}=\pm M|k_w^2-k_{z}^{2}|$ at $k_{x}=k_{y}=0$.
The two bands intersect at $(0,0,\pm k_w)$ (see Fig.~\ref{Fig:band}).

Around the two nodes $(0,0,\pm k_w)$, $H$ reduces to two separate
local models
\begin{equation}\label{Eq:one-node}
H_{\pm}=\mathbf{M}_{\pm}\cdot\sigma,
\end{equation}
$H_{\pm}=\mathbf{M}_{\pm}\cdot\sigma$with $\mathbf{M}_{\pm}=\left(A\widetilde{k}_{x},A\widetilde{k}_{y},\mp2M k_w\widetilde{k}_{z}\right)$
and $(\widetilde{k}_{x},\widetilde{k}_{y},\widetilde{k}_{z})$ the
effective wave vector measured from the Weyl nodes.

\subsection{Berry curvature}

The topological properties in $H$ can be seen from the Berry curvature
\cite{Xiao10rmp}, $\Omega(\mathbf{k})$ = $\nabla_{\mathbf{k}}\times\mathbf{A}(\mathbf{k})$,
where the Berry connection is defined as $\mathbf{A}(\mathbf{k})$
= $i\left\langle u(\mathbf{k})\right|\nabla_{\mathbf{k}}\left|u(\mathbf{k})\right\rangle $.
For example, for the energy eigenstates for the $+$ band $\left|u(\mathbf{k})\right\rangle $
= $[\cos(\Theta/2),\sin(\Theta/2)e^{i\varphi}]$, where $\cos\Theta\equiv\mathcal{M}_{\mathbf{k}}/E_{+}$
and $\tan\varphi\equiv k_{y}/k_{x}$. The three-dimensional Berry
curvature for the two-node model can be expressed as
\begin{eqnarray}\label{Eq:Berry}
\boldsymbol{\Omega}\left(\mathbf{k}\right)=\frac{A^{2}M }{E_{+}^{3}}\left[k_{z}k_{x},k_{z}k_{y},\frac{1}{2}\left(k_{z}^{2}-k_w^{2}-k_{x}^{2}-k_{y}^{2}\right)\right].\nonumber\\
\end{eqnarray}
There exist a pair of singularities at $(0,0,\pm k_w)$
as shown in Fig. {\ref{Fig:band}}. The chirality of a Weyl node
can be found as an integral over the Fermi surface enclosing one Weyl
node $(1/2\pi)\oint\Omega(\mathbf{k})\cdot d\mathbf{S(\mathbf{k})}$,
which yields opposite topological charges $\mp\mathrm{sgn}(M)$
at $\pm k_w$, corresponding to a pair of ``magnetic monopole and
antimonopole'' in momentum space.

\subsection{$k_z$-dependent Chern number}

For a given $k_{z}$, a Chern number
can be well defined as $n_{c}(k_{z})=-(1/2\pi)\iint dk_{x}dk_{y}\Omega(\mathbf{k})\cdot\hat{z}$
to characterize the topological property in the $k_{x}$-$k_{y}$
plane, and \cite{Lu10prb}
\begin{eqnarray}
n_{c}(k_{z})=-\frac{1}{2}[\mathrm{sgn}[M(k_w^2 -k_{z}^{2})]+\mathrm{sgn}(M )].
\end{eqnarray}
The Chern number $n_{c}(k_{z})=-\mathrm{sgn}(M)$
for $-k_w<k_{z}<k_w$, and $n_{c}(k_{z})=0$ otherwise \cite{Yang11prb}.
The nonzero Chern number corresponds to the $k_{z}$-dependent edge
states (known as the Fermi arcs) according to the bulk-boundary correspondence
\cite{Hatsugai93prl}.

\subsection{Fermi arcs}

If there is an open boundary at $y=0$, where the wave function vanishes, the dispersion of the surface states is finally given by \cite{Shen12book,ZhangSB16njp}
\begin{eqnarray}
E_{\text{arc}}(k_x,k_z)=\mathrm{sgn}(M)Ak_x. \label{SurfaceEnergy}
\end{eqnarray}
The corresponding wavefunction is similar to that of topological insulator surface states \cite{Shan11njp,Shen11spin}
\begin{eqnarray}
\Psi_{k_x,k_z}^{\text{arc}}(\mathbf{r}) & = & C e^{ik_{x}x+ik_{z}z} \begin{bmatrix}
\text{sgn}(M) \\ 1
\end{bmatrix} (e^{\lambda_1y}-e^{\lambda_2y}),
\end{eqnarray}
where $C$ is a normalization factor  and $\lambda_{1,2} =A/2|M|\mp\sqrt{(A/2M)^{2}-\Delta_k}$, and $\Delta_k=k_w^2-k_x^2-k_z^2$. There are Fermi arcs in two cases: (i) $\lambda_{1,2}>0$, and (ii) $\lambda_{1,2}= a\mp i b$ with $a,b>0$ (Note that $\lambda_1=\lambda_2$ corresponds to a trivial case).
Also in both cases (i) and (ii), we have $\lambda_1\lambda_2>0$ and henceforth $\Delta_{k}>0$. Therefore the solution of Fermi surface states is restricted inside a circle defined by $k_{x}^{2}+k_{z}^{2}<k_w^{2}$.

The two-node model in
Eq. (\ref{Eq:model}) provides a generic description for Weyl semimetals,
including the band touching, opposite chirality, monopoles of Berry
curvature, topological charges, and Fermi arcs.

\subsection{Monopole charge}\label{Sec:Monopole}

As an example, we use the effective model
\begin{eqnarray}
H &=& v k\cdot \sigma
\end{eqnarray}
to demonstrate the monopole charge hosted at the Weyl nodes. The model is equivalent to Eq. (\ref{Eq:one-node}).
The spinor wave function of the valence band can be found as
\begin{eqnarray}
|u_+(k,\theta,\varphi)\rangle  &=&
\left(
  \begin{array}{c}
    \sin\frac{\theta}{2} \\
    -\cos\frac{\theta}{2}e^{i\varphi} \\
  \end{array}
\right),
\end{eqnarray}
where $\cos\theta\equiv k_z / k$ with $k=\sqrt{k_x^2+k_y^2+k_z^2}$. The Berry connection is defined as
\begin{eqnarray}
\mathbf{A} &\equiv & -i \langle  u_+ \left| \nabla_\mathbf{k} \right| u_+\rangle .
\end{eqnarray}
In polar coordinates, $\nabla_\mathbf{k}=(\partial_k, (1/k)\partial_\theta,(1/k\sin\theta)\partial_\varphi)$, we can find that
\begin{eqnarray}
(A_k,A_\theta,A_\varphi)&=& (0,0,\frac{\cos^2\frac{\theta}{2}}{k\sin\theta}) .
\end{eqnarray}
The Berry curvature can be found as
\begin{eqnarray}
\mathbf{\Omega} &\equiv &  \nabla \times \mathbf{A}
\nonumber\\
&=&
\frac{1}{k\sin\theta} \left[ \frac{\partial(A_\varphi \sin\theta ) }{\partial \theta}     \right]\hat{e}_k  -\frac{1}{k } \frac{\partial(kA_\varphi)}{\partial k} \hat{e}_\theta
\nonumber\\
 &=& -\frac{1}{2k^2} \hat{e}_k.
\end{eqnarray}
The monopole charge is defind as the Berry curvature flux threading a sphere that encloses the origin, and can be found as
\begin{eqnarray}
\mathcal{N}&=&\frac{1}{2\pi}\int_\Sigma d \mathbf{S} \cdot \mathbf{\Omega}\nonumber\\
 &=&  \int_0^{2\pi} d\varphi \int_0^\pi d\theta
\sin\theta k^2 (-\frac{1}{2k^2})\nonumber\\
&=& -1 .
\end{eqnarray}
In the other valley of opposite chirality, the Hamiltonian can be written as $H=-vk\cdot \sigma$, the wave function of the valence band $|u_-\rangle$ can be obtained by letting $\theta\rightarrow \pi/2 -\theta$ and $\varphi\rightarrow \pi +\varphi$ in $|u_+\rangle $, and $|u_-\rangle = (\cos\frac{\theta}{2},\sin\frac{\theta}{2}e^{i\varphi})$. Following the same procedure, we can show that the Berry connection is $ A_\varphi = \sin^2\frac{\theta}{2}/k\sin\theta $, the Berry curvature is $\Omega_k= 1/2k^2 $, and the monopole charge is 1. Thus the total monopole charge is zero for the two-node model, which is consistent with Nielsen-Minomiya's no-go theorem \cite{Nielsen81npb}.

\subsection{Landau bands}

\label{Sec:Landau}
\begin{figure}[tbph]
\centering \includegraphics[width=0.48\textwidth]{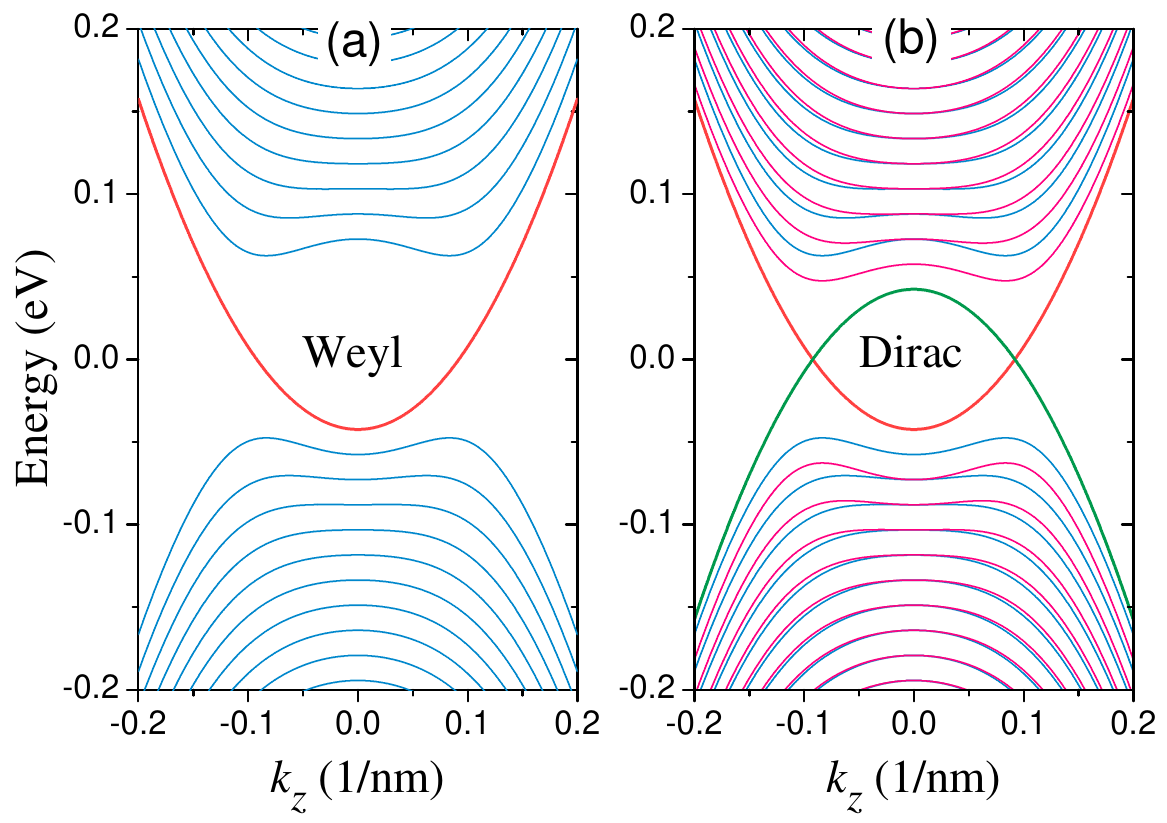}
\caption{The energies of Landau bands of the minimal global model for Weyl
and Dirac semimetals in a magnetic field $B$ applied along the $z$
direction, as functions of the wave vector $k_{z}$. The parameters:
$M_{0}=0.05$ eV, $M_{1}=5$ eV$\cdot$nm$^{2}$, $A=1$ eV$\cdot$nm,
and $B=1$ Tesla. The Zeeman energy is not included. Adapted from \cite{Lu15Weyl-shortrange}.}
\label{Fig:Landau}
\end{figure}

In a magnetic field along the $z$ direction, the energy spectrum
is quantized into a set of 1D Landau bands dispersing with $k_{z}$
[see Fig.~\ref{Fig:Landau} (a)]. We consider a magnetic field
applied along the $z$ direction, $\mathbf{B}=(0,0,B)$, and choose
the Landau gauge in which the vector potential is $\mathbf{A}=(-yB,0,0)$.
The Landau bands can be solved analytically \cite{Shen04prl,Shen04prb,Sakurai-book}.

The eigen
energies are \cite{Lu15Weyl-shortrange}
\begin{eqnarray}\label{Energy-LL}
E_{k_{z}}^{\nu\pm} & = & \omega/2\pm\sqrt{\mathcal{M}_{\nu}^{2}+\nu\eta^{2}},\ \nu\ge1\nonumber \\
E_{k_{z}}^{0} & = & \omega/2-M_{0}+M_{1}k_{z}^{2},\ \ \nu=0,
\end{eqnarray}
where
$\omega=2M /\ell_{B}^{2}$, $\eta=\sqrt{2}A/\ell_{B}$, and the magnetic length
$\ell_{B}=\sqrt{\hbar/e|B|}$.
The Landau energy bands ($\nu$ as band index)
disperse with $k_{z}$, as shown in Fig. \ref{Fig:Landau}. The
eigen states for $\nu\ge1$ are
\begin{eqnarray}\label{LL-nu}
|\nu\ge1,k_{x},k_{z},+\rangle & = & \left[\begin{array}{cc}
\cos\frac{\theta_{k_{z}}^{\nu}}{2}|\nu-1\rangle\\
\sin\frac{\theta_{k_{z}}^{\nu}}{2}|\nu\rangle
\end{array}\right]|k_{x},k_{z}\rangle,\nonumber \\
|\nu\ge1,k_{x},k_{z},-\rangle & = & \left[\begin{array}{cc}
\sin\frac{\theta_{k_{z}}^{\nu}}{2}|\nu-1\rangle\\
-\cos\frac{\theta_{k_{z}}^{\nu}}{2}|\nu\rangle
\end{array}\right]|k_{x},k_{z}\rangle,
\end{eqnarray}
and for $\nu=0$ is
\begin{equation}\label{LL-nu0}
|\nu=0,k_{x},k_{z}\rangle=\left[\begin{array}{cc}
0\\
|0\rangle
\end{array}\right]|k_{x},k_{z}\rangle,
\end{equation}
where $\cos\theta^\nu_{k_z}=\mathcal{M}_{\nu}/\sqrt{\mathcal{M}_{\nu}^{2}+\nu\eta^{2}}$,
and the wave functions $\psi_{\nu,k_{z},k_{x}}(\mathbf{r})=\langle\mathbf{r}|\nu,k_{x},k_{z}\rangle$
are found as
\begin{eqnarray}\label{psi-nukxkz}
\psi_{\nu,k_{z},k_{x}}(\mathbf{r}) & = & \frac{C_{\nu}}{\sqrt{L_{x}L_{z}\ell_{B}}}e^{ik_{z}z}e^{ik_{x}x}e^{-\frac{(y-y_{0})^{2}}{2\ell_{B}^{2}}}\mathcal{H}_{\nu}(\frac{y-y_{0}}{\ell_{B}}),\nonumber \\
\end{eqnarray}
where $C_{\nu}\equiv1/\sqrt{\nu!2^{\nu}\sqrt{\pi}}$, $L_{x}L_{z}$
is area of sample, the guiding center $y_{0}=k_{x}\ell_{B}^{2}$,
$\mathcal{H}_{\nu}$ are the Hermite polynomials. As the dispersions
are not explicit functions of $k_{x}$, the number of different $k_{x}$
represents the Landau degeneracy $N_{L}=1/2\pi\ell_{B}^{2}=eB/h$
in a unit area in the x-y plane.
This set of analytical solutions provides us a good base to study
the transport properties of Weyl fermions.

\subsection{Paramagnetic topological semimetals}

A Weyl semimetal and its time-reversal partner can form a Dirac semimetal or paramagnetic semimetal, whose model can be built by $H(\mathbf{k})$ in Eq. (\ref{Eq:model}) and its time-reversal partner $H^*(-\mathbf{k})$, where the asterisk refers to a complex conjugate. This model can also serve as a building block for Weyl semimetals that respect time-reversal symmetry but break inversion symmetry \cite{Huang15nc,Weng15prx,Lv15prx,Xu15sci-TaAs,Yang15np,ZhangCL16nc,HuangXC15prx,Xu15np-NbAs,Xu16nc-TaP}. For this case,
there is the quantum spin Hall effect, compared to the quantum anomalous Hall effect
in a Weyl semimetal of a single pair of nodes.
A straightforward
extension is as follows \cite{ZhangSB16njp}
\begin{equation}
H_{\text{Dirac}}=A(k_{x}\alpha_{x}+k_{y}\alpha_{y})+M(k_w^{2}-k^{2})\beta,
\end{equation}
where the Dirac matrices are $\alpha_{x}=\sigma_{x}\otimes\sigma_{x}$,
$\alpha_{y}=\sigma_{x}\otimes\sigma_{y}$, $\beta=\sigma_z\otimes\sigma_0$.
It contains four Weyl nodes, which are doubly degenerate. The surface
electrons around the $\hat{\mathbf{z}}$ direction consist of two branches with opposite
spins and opposite effective velocities. The model can also be written into a block-diagonalized form by changing the basis ($1\rightarrow 1$, $4\rightarrow 2$, $2\rightarrow 3$, $3\rightarrow 4$),
\begin{equation}\label{Eq:H-Dirac}
H_{\text{Dirac}}=\left[\begin{array}{cc}
H(\mathbf{k}) & 0\\
0 & H^{*}(-\mathbf{k})
\end{array}\right]+\sigma_{z}\otimes\left[\begin{array}{cc}
\Delta_{s} & 0\\
0 & \Delta_{p}
\end{array}\right].
\end{equation}
In the second term, the $z$-direction Zeeman energy $\Delta_{s/p}=g_{s/p}\mu_{B}B/2$
is also included, where $g_{s/p}$ is the g-factor for the $s/p$
orbital \cite{Wang13prb} and $\mu_{B}$ is the Bohr magneton.

Figure~\ref{Fig:Landau} (b) shows the Landau bands of both $H(\mathbf{k})$
and $H^{*}(-\mathbf{k})$ in the $z$-direction magnetic field. The
Landau bands of the Dirac semimetal can be found in a similar way
as that in Sec. \ref{Sec:Landau}. Now there are two branches of $\nu=0$
bands, with the energy dispersions $E_{k_{z}}^{0\uparrow}=\omega/2+\Delta_{p}-M_{0}+M_{1}k_{z}^{2}$
and $E_{k_{z}}^{0\downarrow}=-\omega/2-\Delta_{s}+M_{0}-M_{1}k_{z}^{2}$
for $H(\mathbf{k})$ and $H^{*}(-\mathbf{k})$, respectively. They
intersect at $k_{z}=\pm\sqrt{[M_{0}-(\omega+\Delta_{s}+\Delta_{p})/2]/M_{1}}$
and energy $(\Delta_{p}-\Delta_{s})/2$, and with opposite Fermi velocities
near the points.

\subsection{Double-Weyl semimetal}

Each Weyl node in a Weyl semimetal hosts a monopole charge of 1 or -1. In a doulbe-Weyl semimetal, the monopole charge is 2 or -2 \cite{Xu11prl,Fang12prl,Guan15prl,HuangSM16pnas-SrSi2}. For a single valley of both single- and double-Weyl semimetals, the minimal model can be written as
\begin{equation}\label{hamiltonian}
H=
\left[
  \begin{array}{cc}
    \chi v_z \hbar k_z & v_{\Vert}(\hbar k_{+})^\mathcal{N} \\
    v_{\Vert} (\hbar k_{-})^\mathcal{N} & -\chi v_z \hbar k_z \\
  \end{array}
\right],
\end{equation}
where $k_{\pm}=k_x\pm i k_y$, $\chi=\pm 1$ is the valley index, $v_z$ and $v_{||}$ are parameters and assumed to be constants, and momentum $\mathbf{k}$ is measured from the Weyl nodes.
Here, $\mathcal{N}=1,2$ correspond to single- and double-Weyl semimetal respectively.
The model has a conduction band and a valence band, with the dispersions given by $\pm E_\mathbf{k}$ and $ E_\mathbf{k}= \sqrt{v_z^2 \hbar^2 k_z^2+v_{||}^2 (\hbar^2 k_x^2+\hbar^2 k_y^2)^{\mathcal{N}}}$.
Without loss of generality, we assume that the chemical potential is slightly above the Weyl nodes and the electronic transport is contributed mainly by the
conduction bands throughout the paper. The eigenstate of the conduction band at valley $\chi=+$ is given by
\begin{eqnarray}\label{eigen-states}
 |\mathbf{k}\rangle =
 \left[\begin{array}{c}
 \cos (\theta/2)\\
 \sin (\theta/2)\exp(-i \mathcal{N} \varphi)\\
 \end{array}\right],
\end{eqnarray}
where
$\cos \theta\equiv v_z k_z/E_\mathbf{k} $, and $\tan \varphi\equiv k_y/k_x$.
The eigenstate of the conduction band around valley $\chi=-$ can be found by replacing $\cos(\theta/2)\rightarrow \sin(\theta/2)$ and $\sin(\theta/2)\rightarrow -\cos(\theta/2)$ in Eq. (\ref{eigen-states}).
The monopole charge can be found by integrating the Berry curvature over an arbitrary Fermi sphere $\Sigma$ that encloses the Weyl node,
\begin{equation}\label{charge}
\frac{1}{2\pi} \int_{\Sigma} d \mathbf{S}\cdot \mathbf{\Omega}=\pm \mathcal{N},
\end{equation}
with $\pm$ for the $\pm$ valleys, the Berry curvature \cite{Xiao10rmp} $\mathbf{\Omega}=\nabla \times \mathbf{A}$, and $\mathbf{A}=(A_{\theta},A_{\varphi})$ is the Berry connection given by
$A_{\theta}=\langle \mathbf{k}| i \partial_{\theta}|\mathbf{k} \rangle=0$ and $ A_{\varphi}=\langle \mathbf{k}|i \partial_{\varphi}|\mathbf{k} \rangle =\mathcal{N} \sin^2 (\theta/2)$.

\section{Near zero field: Weak anti-localization}\label{Sec:WAL}

Weak anti-localization is a transport phenomenon in disordered metals \cite{Lee85rmp}. At low temperatures, when the mean free path is much shorter than the system size and phase coherence length, electrons suffer from scattering but can maintain their phase coherence. In this quantum diffusive regime, the quantum interference between time-reversed scattering loops can give rise to a correction to the conductivity. If the quantum interference correction is positive, it gives a weak anti-localization correction to the conductivity. Because this correction requires time reversal symmetry, it can be suppressed by applying a magnetic field, leading to a negative magnetoconductivity, or positive magnetoresistivity, as the signature for the weak anti-localization. The weak anti-localization has been widely observed in topological topological semimetals, including Bi$_{0.97}$Sb$_{0.03}$
,\cite{Kim13prl,Kim14prb} ZrTe$_5$,\cite{Li16np}, Na$_3$Bi \cite{Xiong15sci}, Cd$_3$As$_2$ \cite{LiCZ15nc,LiH16nc}, TaAs \cite{HuangXC15prx,ZhangCL16nc}, etc.

\subsection{Symmetry argument}

In contrast, the quantum interference can be negative, leading to the weak localization effect and totally opposite temperate and magnetic dependencies of conductivity.
Whether one has weak localization or weak anti-localization
depends on the symmetry (see Table \ref{Tab:symmetry}). According to the classification of the ensembles of random matrix \cite{Dyson62jmp}, there are three symmetry classes. If a system has time-reversal symmetry but no spin-rotational symmetry, it is in the symplectic class, in which the weak anti-localization is expected \cite{Hikami80ptp}.
Remember that one of the low-energy descriptions of Weyl fermions in semimetals is
$H=\pm\hbar v_{F}\boldsymbol{\sigma}\cdot\mathbf{k}$, which respects time-reversal symmetry not spin rotational symmetry. Therefore, a single valley of Weyl fermions has the symplectic symmetry and the weak anti-localization. Moreover, we find the Berry phase can also explain the weak localization in Weyl semimetals \cite{Dai16prbrc}, which we discuss later.

\begin{table}[htbp]
\caption{The relation between the symmetry classes (orthogonal, symplectic, and unitary) \cite{Dyson62jmp} and weak localization (WL) and anti-localization (WAL) \cite{Hikami80ptp}. Adapted from Ref. \cite{Lu16cpb}.}
\label{Tab:symmetry}
\centering
\begin{tabular}{cccc}
\hline
&   Orthogonal &  Symplectic    & Unitary  \\
  \hline
Time-reversal   &  $\surd$ & $\surd$ & $\times $  \\
Spin-rotational & $\surd$ & $\times$ & $\times$   \\
WL/WAL  & WL & WAL &  $\times$    \\
\hline
\end{tabular}
\end{table}

%For Dirac fermions, the weak anti-localization has an alternative understanding based on the Berry phase argument.
%The Berry phase is a geometric phase collected in an adiabatic cyclic process.\cite{Pancharatnam56,Berry84}
%Since the studies on carbon nanotubes, it has been found that massless Dirac fermions can collect a $\pi$ Berry phase after circulating around the Fermi surface.\cite{Shon98jpsj} The $\pi$ Berry phase induces a destructive quantum interference between time-reversed loops formed by scattering trajectories. The destructive interference can suppress backscattering of electrons, then the conductivity is enhanced with decreasing temperature, as decoherence mechanisms are suppressed at low temperatures.\cite{Suzuura02prl,McCann06prl}

\subsection{Feynman diagram calculations}

\begin{figure}[tbph]
\centering \includegraphics[width=0.45\textwidth]{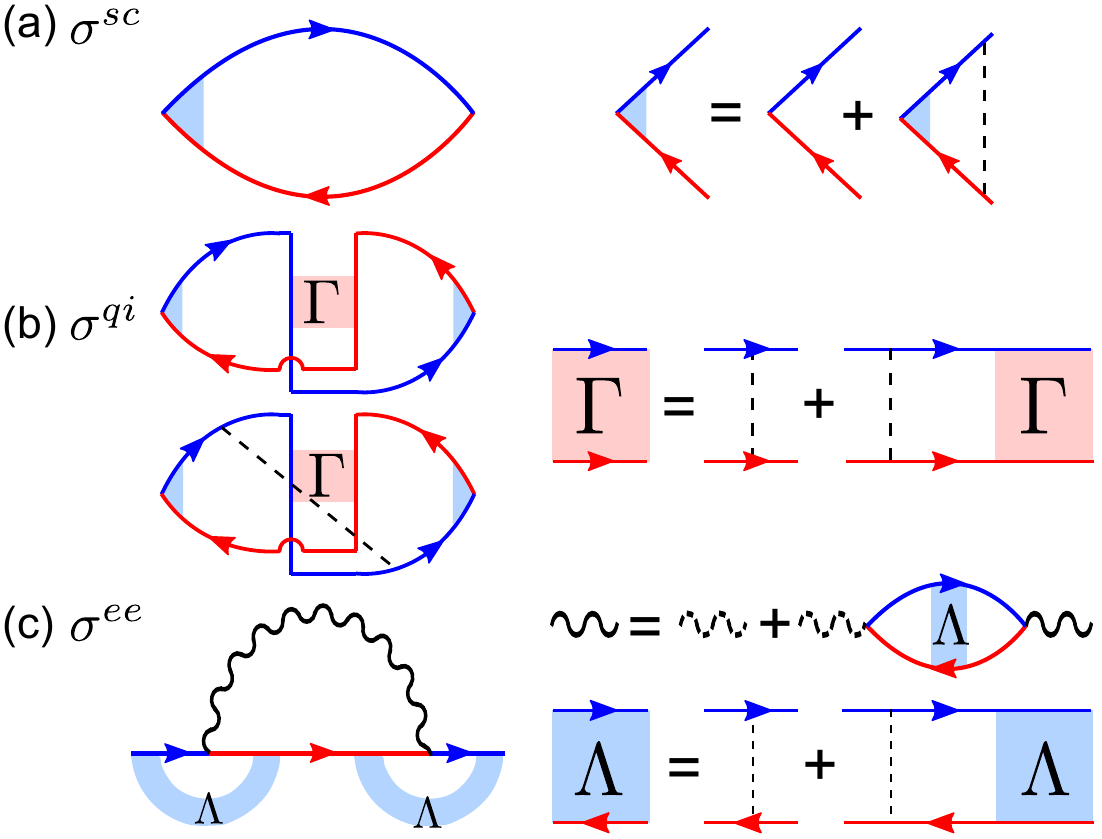}
\protect\caption{The Feynman diagrams \cite{McCann06prl,Altshuler80prl,Fukuyama80jpsj,Lee85rmp,Lu11prl,Shan12prb,Lu14prl}
for the conductivity of 3D Weyl semimetals, in the presence of disorder
(dashed lines) and electron-electron interaction (wavy lines). The
arrow lines are for Green's functions. Adapted from Ref. \cite{Lu15Weyl-Localization}.}
\label{Fig:Feynman-Diagram}
\end{figure}

One of the theoretical approaches to study the weak localization and anti-localization is the Feynman diagram techniques. Figure \ref{Fig:Feynman-Diagram} summarizes the Feynman diagrams used to study the weak localization and anti-localization arising from the quantum interference and interaction \cite{Lu15Weyl-Localization}. It is based on the linear response theory of the conductivity, with disorder and interaction taken as perturbations. In the formulism, there are three main contributions to the conductivity. The leading order is the semiclassical Drude conductivity [Fig. \ref{Fig:Feynman-Diagram}(a)], then the quantum interference correction [Fig. \ref{Fig:Feynman-Diagram}(b)] and interaction correction (Altshuler-Aronov effect) [Fig. \ref{Fig:Feynman-Diagram}(d)].

\begin{figure}[tbph]
\centering
\includegraphics[width=0.49\textwidth]{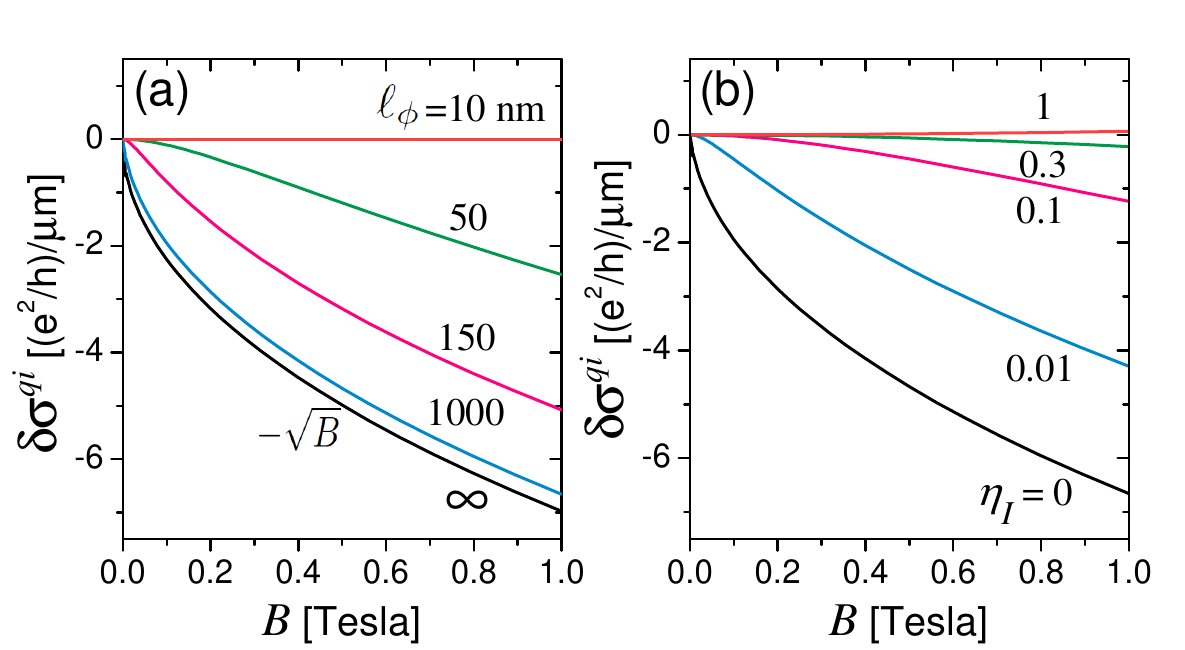}
\includegraphics[width=0.48\textwidth]{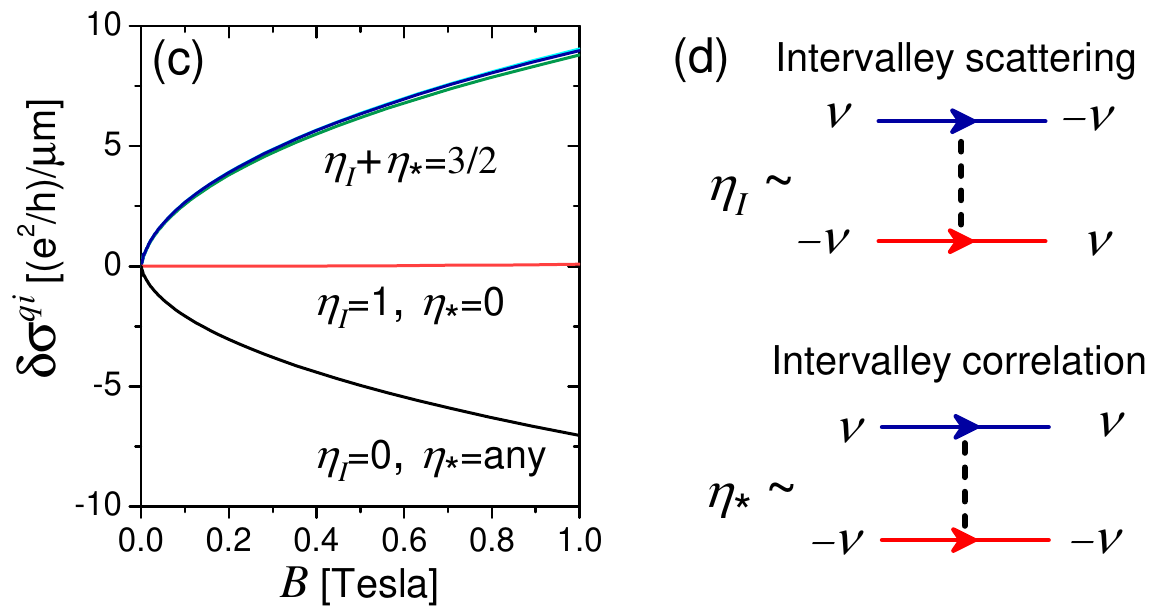}
\protect\caption{The magnetoconductivity $\delta\sigma^{qi}(B)$ for different phase
coherence length $\ell_{\phi}$ at $\eta_{I}=\eta_{*}=0$ (a), for
different $\eta_{I}$ at $\eta_{*}=0$ (b), and for different $\eta_{I}$
at finite $\eta_{*}$ (c). The magnetic field $B$ is applied along arbitrary
directions. Parameters: $\ell=10$ nm and $\ell_{\phi}=1000$
nm in (b) and (c). (d) The diagrams show the difference between $\eta_{I}$
and $\eta_{*}$, with $\eta_{I}$ related to the intervalley scattering
and $\eta_{*}$ measuring the intervalley correlation of intravalley
scattering. The dashed lines represent the correlation of two scattering
processes. $\nu=\pm$ is the valley index. Adapted from Ref. \cite{Lu15Weyl-Localization}.}
\label{Fig:Weyl-MC}
\end{figure}

We calculate the magnetoconductivity arising from the quantum interference $\delta\sigma^{qi}$, as shown in Fig. \ref{Fig:Weyl-MC}. As $B\rightarrow0$, $\delta\sigma^{qi}$ is
proportional to $-\sqrt{B}$ for $\ell_{\phi}\gg\ell_{B}$ or at low
temperatures, and $\delta\sigma^{qi}\propto-B^{2}$ for $\ell_{\phi}\ll\ell_{B}$ or at high temperatures.
$\ell_{B}$ can be evaluated approximately as 12.8 nm$/\sqrt{B}$
with $B$ in Tesla. Usually below the liquid helium temperature, $\ell_{\phi}$
can be as long as hundreds of nanometers to one micrometer, much longer than
$\ell_{B}$ which is tens of nanometers between 0.1 and 1 Tesla. Therefore,
the $-\sqrt{B}$ magnetoconductivity at low temperatures and small
fields serves as a signature for the weak anti-localization of 3D Weyl fermions.
Fig. \ref{Fig:Weyl-MC}(a) shows $\delta\sigma^{qi}(B)$ of two valleys
of Weyl fermions in the absence of intervalley scattering. For long $\ell_{\phi}$,
$\delta\sigma^{qi}(B)$ is negative and proportional to $\sqrt{B}$,
showing the signature of the weak anti-localization of 3D Weyl fermions. This
$-\sqrt{B}$ dependence agrees well with the experiment,\cite{Kim13prl,Kim14prb}
and we emphasize that it is obtained from a complete diagram calculation
with only two parameters $\ell$ and $\ell_{\phi}$ of physical meanings.
As $\ell_{\phi}$ becomes shorter, a change from $-\sqrt{B}$ to $-B^{2}$
is evident, and eventually $\delta^{qi}(B)$ vanishes at $\ell_{\phi}=\ell$ as the
system is no longer in the quantum interference regime and enters the semiclassical diffusion regime.

\subsection{Weak localization of double-Weyl semimetal}

\begin{figure}[tb]
\centering
\includegraphics[width=0.5\textwidth]{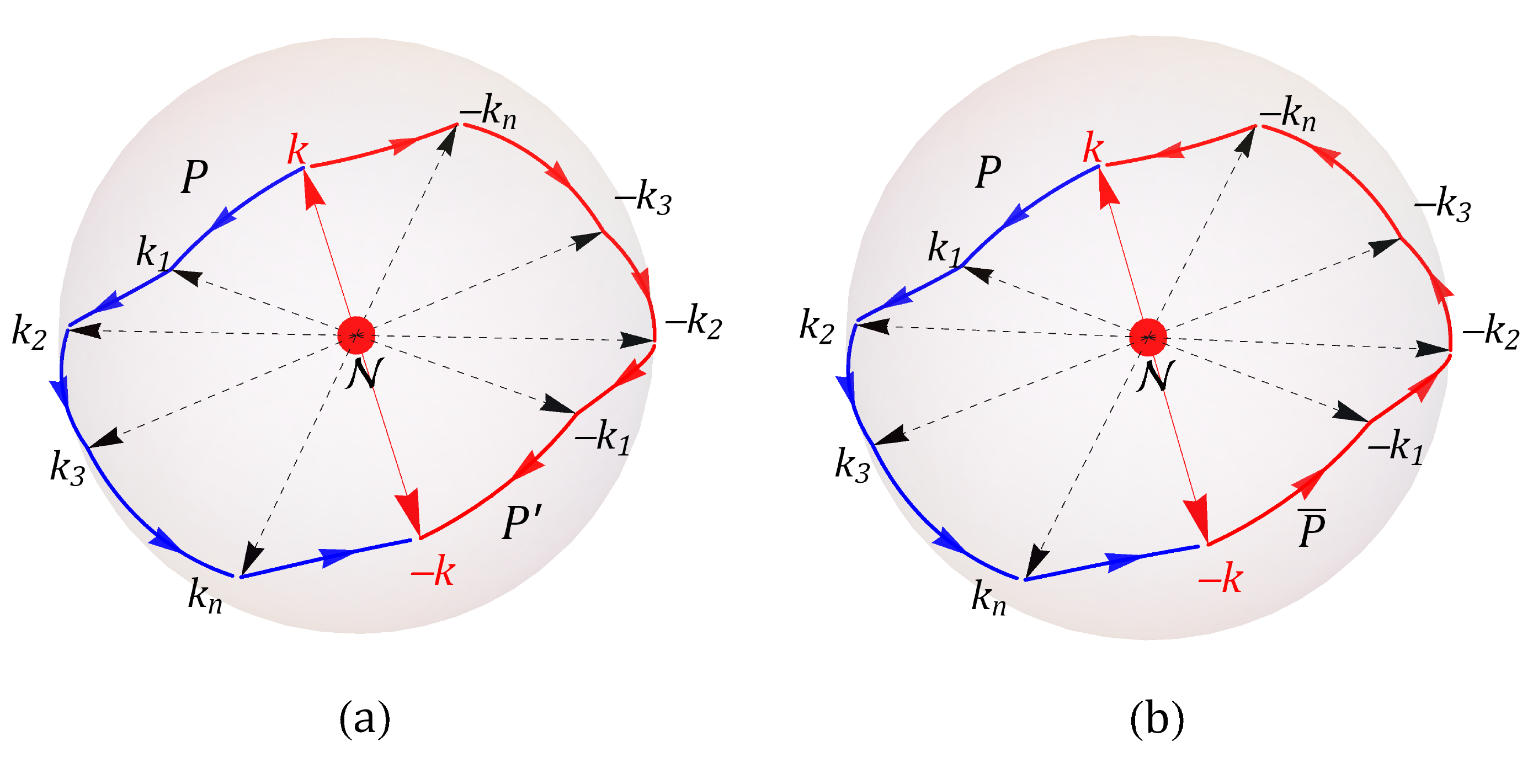}
\protect\caption{The Fermi sphere in momentum space for a three-dimensional topological semimetal, where the dot located at the origin represents a monopole charge of $\mathcal{N}$.
(a) $P$ denotes a generic backscattering from the wave vector $\mathbf{k}$ to $-\mathbf{k}$ via intermediate states labeled as ($\mathbf{k}_1,\mathbf{k}_2,...,\mathbf{k}_n$). $P'$ stands for the time-reversal counterpart of $P$.
(b) The phase difference between $P$ and $P'$ is equivalent to the Berry phase circulating around the loop $\mathcal{C}=P+\bar{P}$. Adapted from Ref. \cite{Dai16prbrc}.}
\label{Fig:loop}
\end{figure}

We focus on the Fermi sphere in one valley of a topological semimetal. For each path [labelled as $P$ in Fig.~\ref{Fig:loop}(a)] connecting successive intermediate states of the backscattering from $\mathbf{k}$ to $-\mathbf{k}$ on the Fermi sphere, which encompasses the monopole charge at the origin, there exists a corresponding time-reversal counterpart $P'$. The quantum interference is determined by the phase difference between the two time-reversed paths $P$ and $P'$, which is equivalent to the Berry phase accumulated along the loop formed by $P$ together with $\bar P\equiv -P'$, namely the corresponding path from $-\mathbf{k}$ to $\mathbf{k}$, as shown in Fig. \ref{Fig:loop}(b).

The quantum interference correction then depends on the geometric phase, i.e., the Berry phase \cite{Xiao10rmp,Pancharatnam56,Berry84,Shon98jpsj,Suzuura02prl,McCann06prl}, collected by electrons after circulating the loop $\mathcal{C}\equiv P+\bar P$.
The Berry phase can be found by a loop integral of the Berry connection around $\mathcal{C}$.
Remarkably, this Berry phase depends only on the monopole charge, but {\it not} on the specific shape of the loop \cite{Dai16prbrc}
\begin{equation}\label{charge_berry}
\gamma =\oint_{\mathcal{C}} d \mathbf{\ell} \cdot \mathbf{A}= \pi \mathcal{N}.
\end{equation}
For double-Weyl semimetals, the monopole charge $\mathcal{N}=2$ and the Berry phase is then $2\pi$. With the $ 2\pi$ Berry phase, the time-reversed scattering loops interfere constructively, leading to the weak localization effect.
However, for single-Weyl semimetals, the monopole charge is $\mathcal{N}= 1$ and the Berry phase is $\pi$, which gives rise to the weak anti-localization effect. As the Berry phase is a consequence of the Berry curvature field generated by the monopole charge, we therefore establish a robust connection between the weak (anti)localization effect with the parity of monopole charge $\mathcal{N}$.
The Berry phase argument is consistent with the symmetry classification\cite{Altland97prb},
the single-Weyl	semimetals belong to the symplectic class with a weak anti-localization correction,
while double-Weyl semimetals correspond to the orthogonal class with a weak localization correction.

\begin{figure}[tb]
\centering
\includegraphics[width=0.25\textwidth]{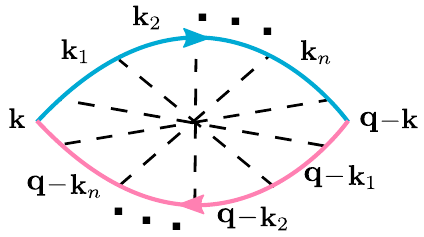}
\protect\caption{The maximally crossed Feynman diagram that describes the quantum interference between the time-reversed scattering trajectories in Fig. \ref{Fig:loop} as $\mathbf{q}\rightarrow 0$. The arrowed solid and dashed lines denote the Green functions and impurity scattering, respectively. This kind of diagrams can give the quantum interference correction to the conductivity \cite{Hikami80ptp,Shon98jpsj,McCann06prl}.
A negative (positive) correction corresponds to the weak (anti)localization effect, with the sign sensitive to the parity of the monopole charge. Adapted from Ref. \cite{Dai16prbrc}.}
\label{Fig:diagram}
\end{figure}

We now verify the above argument of quantum interference correction to conductivity in Weyl semimetals by the standard Feynman diagram calculations.
The correction can be evaluated by calculating the maximally crossed diagrams, one of which is shown in Fig.~\ref{Fig:diagram}. In this diagram, the segments of the arrow lines represent the intermediate states in the backscattering, and the dashed lines represent the correlation between the time-reversed scattering processes. The core calculation of the maximally crossed diagrams can be formulated into the particle-particle correlation, known as the cooperon.
The cooperon of the double-Weyl semimetal is found to be \cite{Dai16prbrc}
\begin{equation}\label{Eq:cooperon2}
\Gamma _{\mathbf{k}_1,\mathbf{k}_2}\approx \frac{\hbar }{2\pi N_F \tau ^2}\frac{e^{i2(\varphi_2-\varphi_1)}}{ D_1\left(q_x^2+q_y^2\right)+D_2 q_z^2},
\end{equation}
where $\mathbf{q}=\mathbf{k}_1+\mathbf{k}_2$ is the cooperon wave vector, $\mathbf{k}_1$ and $\mathbf{k}_2$ are the wave vectors of incoming and outgoing states, respectively,
$\varphi_1$ and $\varphi_2$ are the azimuth angles of corresponding wave vectors,
$D_1= 8 \tau E_F v_{\Vert} /3 \pi $ and $D_2= \tau v_z^2$ are the diffusion coefficients,
$N_F$ is the density of states, and $\tau$ is the transport time.
In contrast, the cooperon of the single-Weyl semimetal is known to take the form \cite{Lu15Weyl-Localization}
\begin{equation}\label{Eq:cooperon1}
\Gamma_{\mathbf{k}_1,\mathbf{k}_2}
\approx \frac{\hbar}{2\pi N_F \tau^2} \frac{1}{ D q^2}e^{i(\varphi_2-\varphi_1)},
\end{equation}
where the diffusion coefficient $D=v^2_F \tau/2$
(We only give the result for isotropic single-Weyl semimetals with $v_F=v_z=v_{\Vert}$; this simplification does not change any qualitative results with respect to quantum interference correction). Note the main difference between Eqs. \eqref{Eq:cooperon2} and \eqref{Eq:cooperon1} lies in the phase factor involving $\varphi_2-\varphi_1$, which originates from different eigenstates of Weyl semimetals with different monopole charges.

As $\mathbf{q}\to 0$, i.e., $\mathbf{k}_1=-\mathbf{k}_2$, the cooperon becomes divergent and becomes the most dominant contribution to the backscattering. In this limit, $\varphi_2= \varphi_1+ \pi$
(We have carried out a coordinate transformation in deriving these results, where $\hbar k_x=\sqrt{k \sin \theta }\cos \varphi, \hbar k_y=\sqrt{k \sin \theta }\sin \varphi, 2\hbar m v k_z =k \cos \theta $, $-\mathbf{k}$ is obtained by setting $\varphi \to \varphi+ \pi$ and $\theta\to \pi-\theta$).
Then, for the double-Weyl semimetal,
\begin{eqnarray}\label{Eq:coop2}
\Gamma_{\mathbf{k},\mathbf{q}-\mathbf{k}}&\approx &+\frac{\hbar }{2\pi N_F \tau ^2}\frac{1}{ D_1\left(q_x^2+q_y^2\right)+D_2 q_z^2},
\end{eqnarray}
and for the single-Weyl semimetal,
\begin{eqnarray}\label{Eq:coop1}
\Gamma_{\mathbf{k},\mathbf{q}-\mathbf{k}}
&\approx &-\frac{\hbar}{2\pi N_F \tau^2} \frac{1}{ D q^2}.	
\end{eqnarray}
Note the different signs in Eqs. \eqref{Eq:coop2} and \eqref{Eq:coop1}, which correspond to the WL and WAL effects, respectively.
This is a direct consequence of different phase factors in the wavefunctions, generated by different monopole charges in double- and single-Weyl semimetals.
In other words, a connection is therefore firmly established between the \emph{parity} of monopole charge $\mathcal{N}$ and the sign of the quantum interference correction, with odd and even parity giving rise to WAL and WL, respectively.

The weak localization effect can give rise to a positive magnetoconductivity as another signature of the weak localization in double-Weyl semimetals.
The magnetoconductivity is anisotropic, depending on whether the field is along the $z$ direction or in the $x-y$ plane. The magnetoconductivity is defined as $\delta\sigma^\textrm{qi}_{zz}(B)=\sigma^\textrm{qi}_{zz}(B)-\sigma^\textrm{qi}_{zz}(0)$. %In both directions, the magnetoconductivity shows similar the functional relations.
In the limit of $\ell_{\phi}\gg \ell_B \gg \ell_z$, which can be approached at low temperatures, the magnetoconductivity $\delta \sigma^\textrm{qi}_{zz}(B) \propto \sqrt{B}$. In the limit of $\ell_B\gg \ell_\phi$ and $\ell_B\gg \ell_z$, $\delta \sigma^\textrm{qi}_{zz}(B) \propto B^2$.

\subsection{Magnetoconductivity formula for WAL/WL in 3D}

Based on our theoretical results in Refs. \cite{Lu15Weyl-Localization} and \cite{Dai16prbrc}, we proposed a formula to fit the magnetoconductivity arising from the weak (anti-)localization in three dimensions,
\begin{eqnarray}\label{fitting-formula}
\delta\sigma^\textrm{qi}_{zz} &=& C^\textrm{qi}_1\frac{B^2\sqrt{B}}{B_c^2+B^2}+C^\textrm{qi}_2\frac{B_c^2 B^2}{B_c^2+B^2},
\end{eqnarray}
where the fitting parameters $C^\textrm{qi}_{1}$ and $C^\textrm{qi}_{2}$ are positive for weak localization and negative for weak anti-localization. The critical field $B_c$ is related to the phase coherence length $\ell_\phi$ according to $B_c\sim \hbar/e\ell_\phi^2$. Empirically, the phase coherence length becomes longer with decreasing temperature and can be written as $\ell_\phi \sim T^{-p/2}$; then $B_c \sim   T^p$, where $p$ is positive and determined by decoherence mechanisms such as electron-electron interaction ($p=3/2$) or electron-phonon interaction ($p=3$). At high temperatures, $\ell_\phi\rightarrow 0$; thus, $B_c \rightarrow \infty$ and we have $\delta\sigma^\textrm{qi}_{zz}\propto B^2$. At low temperatures, $\ell_\phi\rightarrow \infty$; then $B_c=0$ and we have $\delta\sigma^\textrm{qi}_{zz}\propto\sqrt{B}$. The formula has been applied in the experiment on TaAs, and by fitting the magnetoconductivity, we find that $p\approx 1.5$ \cite{ZhangCL16nc}.

\begin{figure}[tbph]
\centering \includegraphics[width=0.48\textwidth]{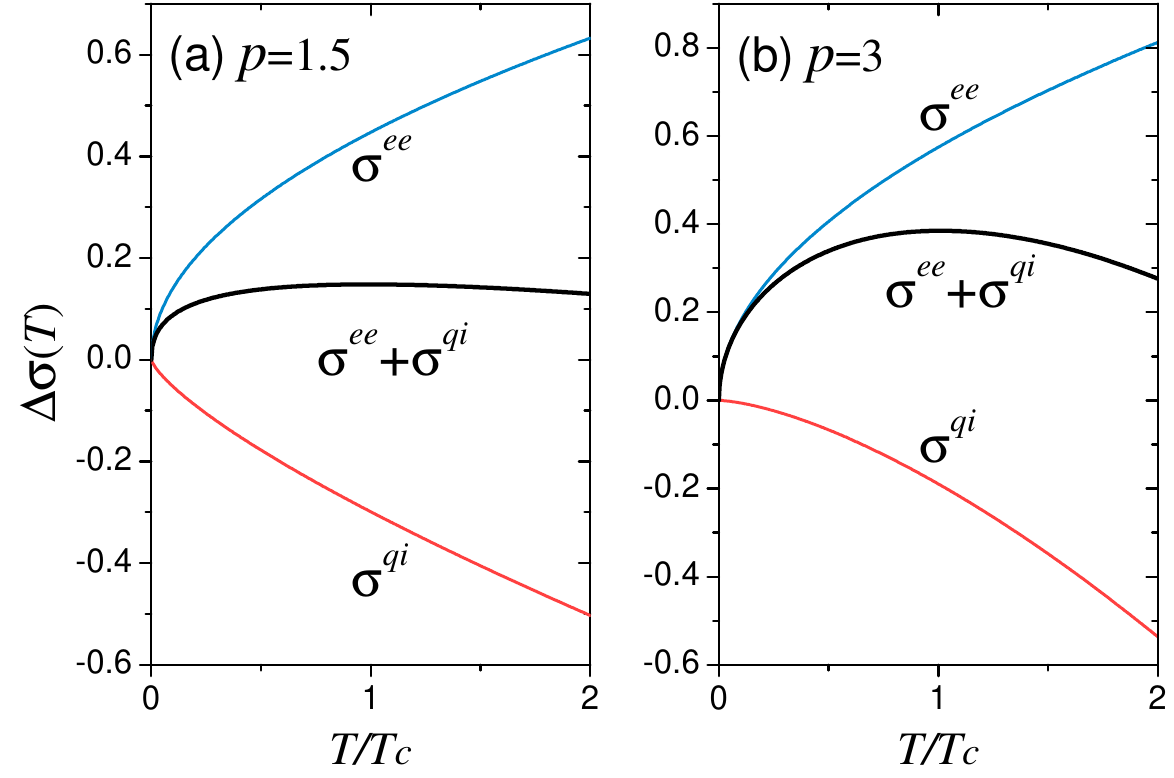}
\protect\caption{A schematic demonstration of the change of conductivity $\Delta\sigma$
as a function of temperature $T$. We choose $c_{ee}=c_{qi}$. $T_{c}$
is the critical temperature below which the conductivity drops with
decreasing temperature. Adapted from Ref. \cite{Lu15Weyl-Localization}.}
\label{Fig:Weyl-sigma-T}
\end{figure}

\subsection{Localization induced by interaction and inter-valley effects}
In the presence of the interaction, we find that the change of conductivity
with temperature for one valley of Weyl fermions can be summarized
as \begin{eqnarray}
\Delta\sigma(T)=c_{ee}T^{1/2}-c_{qi}T^{p/2},
\end{eqnarray}
where both $c_{ee}$ and $c_{qi}$ are positive parameters. This describes a competition between the interaction-induced
weak localization and interference-induced weak anti-localization, as shown in
Fig. \ref{Fig:Weyl-sigma-T} schematically. At higher temperatures, the
conductivity increases with decreasing temperature, showing a weak anti-localization
behavior. Below a critical temperature $T_{c}$, the conductivity
starts to drop with decreasing temperature, exhibiting a localization
tendency. The critical
temperature can be found as $T_{c}=\left(c_{ee}/p\cdot c_{qi}\right)^{2/(p-1)}$.
Because $c_{ee},c_{qi}>0$, this means as long as $p>1$, there is
always a critical temperature, below which the conductivity drops
with decreasing temperature. For known decoherence mechanisms in 3D,
$p$ is always greater than 1 \cite{Lee85rmp}. With a set of typical parameters, we find that $T_{c}\approx0.4\sim10^{6}$
K \cite{Lu15Weyl-Localization}.

We find that the intervalley scattering and correlation can also lead to the weak localization. Two dimensionless parameters are defined for the inter- and
intravalley scattering: $\eta_{*}\propto\langle U_{\mathbf{k},\mathbf{k}'}^{++}U_{\mathbf{k}',\mathbf{k}}^{--}\rangle$
measuring the correlation between intravalley scattering and $\eta_{I}\propto\langle U_{\mathbf{k},\mathbf{k}'}^{+-}U_{\mathbf{k}',\mathbf{k}}^{-+}\rangle$
measuring the weight of intervalley scattering , where $U_{\mathbf{k},\mathbf{k}'}^{\nu,\nu'}$
is the scattering matrix element. Figure \ref{Fig:Weyl-MC}(d)
schematically shows the difference between $\eta_{*}$ and $\eta_{I}$.
As shown in Fig. \ref{Fig:Weyl-MC}(b),
with increasing $\eta_{I}$, the negative $\delta\sigma^{qi}$ is
suppressed, where $\eta_{I}\rightarrow1$ means strong intervalley
scattering while $\eta_{I}\rightarrow0$ means vanishing intervalley
scattering. Furthermore, Fig. \ref{Fig:Weyl-MC}(c) shows that the magnetoconductivity
can turn to positive when $\eta_{I}+\eta_{*}=3/2$.
The positive $\delta\sigma^{qi}(B)$ in Fig. \ref{Fig:Weyl-MC}(c)
corresponds to a suppressed $\sigma^{qi}$ with decreasing temperature,
i.e., a localization tendency. This localization is attributed to the strong intervalley coupling which recovers spin-rotational symmetry (now the spin space is complete for a given momentum), then the system goes to the orthogonal class \cite{Dyson62jmp,Hikami80ptp,Suzuura02prl}. Therefore, we show that the combination
of strong intervalley scattering and correlation will strengthen the
localization tendency in disordered Weyl semimetals. The metal-insulator phase transition is also found numerically \cite{ChenCZ15prl}.

\section{Weak magnetic fields: Negative magnetoresistance}\label{Sec:NMR}

In a topological semimetal, paired Weyl nodes carry opposite chirality and paired monopoles and anti-monopoles of Berry curvature in momentum space\cite{Volovik03book} [see \ref{Fig:band}(b)]. The nontrivial Berry curvature can couple an external magnetic field to the velocity of electrons, leading to a chiral current that is linearly proportional to the field. The correlation of chiral currents further contributes to an extra conductivity that quadratically grows with increasing magnetic field, in a magnetic field and an electric field applied parallel to each other \cite{Son13prb,Burkov14prl-chiral}. This $B^2$ positive conductivity in weak parallel magnetic fields, or negative magnetoresistance (negative MR), is rare in non-ferromagnetic materials, thus can serve as one of the transport signatures of the topological semimetals. More importantly, because of its relation to the chiral charge pumping between paired Weyl nodes, the negative magnetoresistance is also believed to be a signature of the chiral anomaly \cite{Adler69pr,Bell69Jackiw,Nielsen81npb}. The negative magnetoresistance has been observed in topological insulator thin films \cite{Wang12nr} and many topological semimetals, including BiSb alloy\cite{Kim13prl,Kim14prb}, ZrTe$_5$ \cite{Li16np}, TaAs \cite{ZhangCL16nc,HuangXC15prx}, Na$_3$Bi \cite{Xiong15sci}, Cd$_3$As$_2$ \cite{LiCZ15nc,ZhangC15arXiv,LiH16nc}, TaP \cite{Arnold16nc}, NbAs \cite{YangXJ15arXiv,YangXJ15arXiv-NbAs}, and HfTe$_5$ \cite{WangHC16prb}.

To understand the negative magnetoresistance, we start with the semiclassical equation of motion proposed by Niu and his colleagues \cite{Chang95prl,Sundaram99prb,Xiao10rmp,ZhouJH13cpl}
\begin{eqnarray}\label{Eq:Niu-eqn}
\mathbf{\dot{r}} &=& \mathbf{v}+\mathbf{\dot{k}}\times \mathbf {\Omega}_\mathbf{k} \nonumber\\
\hbar\mathbf{\dot{k}} &=& e\mathbf{E}+e\mathbf{\dot{r}}\times \mathbf{B}
\end{eqnarray}
where $\mathbf{v}=\ \partial \epsilon_\mathbf{k}/\hbar \partial \mathbf{k} $. The second term in the first equation indicates that an electron can acquire an anomalous velocity
proportional to the Berry curvature of the band in the presence of an
electric field. This anomalous velocity is responsible for a number of
transport phenomena.

Iterating $\mathbf{k}$ and $\mathbf{r}$ in the equations, using
$(\mathbf{a}\times \mathbf{b})\times \mathbf{c}=(\mathbf{a}\cdot \mathbf{c} )\mathbf{b}-(\mathbf{b}\cdot \mathbf{c})\mathbf{a}
$, and $(\mathbf{\dot{a}}\times \mathbf{b})\cdot \mathbf{b}=0$, we arrive at \cite{Son13prb}
\begin{eqnarray}\label{Eq:Niu-eqn2}
\mathbf{\dot{r}} &=& \left(1+\frac{e}{\hbar} \mathbf{B}\cdot \mathbf {\Omega}_\mathbf{k}\right)^{-1} \left[\mathbf{v}+ \frac{e}{\hbar} \mathbf{E}\times \mathbf {\Omega}_\mathbf{k}+\frac{e}{\hbar} ( \mathbf {\Omega}_\mathbf{k}\cdot\mathbf{v})\mathbf{B}  \right], \nonumber\\
\hbar\mathbf{\dot{k}} &=& (1+\frac{1}{\hbar} \mathbf{B}\cdot\mathbf {\Omega}_\mathbf{k})^{-1}\left[e\mathbf{E}+ e\mathbf{v}\times \mathbf{B}+\frac{e^2}{\hbar}( \mathbf{E}\cdot \mathbf{B})\mathbf {\Omega}_\mathbf{k}   \right],\nonumber\\
\end{eqnarray}
where $\mathbf{E}\times \mathbf {\Omega}_\mathbf{k}$ gives AHE \cite{Goswami13prb,Zyuzin12prb},
$(\mathbf {\Omega}_\mathbf{k}\cdot \mathbf{v})\mathbf{B}$ gives the chiral magnetic effect \cite{Chang15prb}, and
$(\mathbf{E}\cdot \mathbf{B})\mathbf {\Omega}_\mathbf{k}$ is the source of the negative magnetoresistance \cite{Son13prb,Burkov14prl-chiral}.

Now we give an argument for the negative magnetoresistance. The argument is similar to the calculation by Yip \cite{Yip15arXiv}. In the framework of linear response theory, $\mathbf{E}=0$, the velocity in small $B$ fields reduces to
\begin{eqnarray}
\mathbf{\dot{r}}
 &= & \mathbf{v} +\frac{e}{\hbar}(\mathbf {\Omega}_\mathbf{k}\cdot \mathbf{v})\mathbf{B},
\end{eqnarray}
where we have considered the correction of the density of states by the Berry curvature.
The second term represents the anomalous velocity induced by the finite Berry curvature and is proportional to the magnetic field. Because the conductivity is a current-current correlation [see Fig. \ref{Fig:Feynman-Diagram}(a)], the linear-$B$ dependence in the velocity (note that current is charge times velocity) leads to the quadratic-$B$ dependence in the conductivity. In Sec. \ref{Sec:Monopole}, we have shown that the Berry curvature is proportional to $1/k^2$. Considering there are $\Omega^2$ and a $k^2$ in the 3D integral of the conductivity formula, eventually, the anomalous conductivity part should be inversely proportional to the Fermi wave vector and proportional to $B^2$, that is
\begin{eqnarray}
\delta \sigma (B) \propto \frac{B^2}{k_F^2}¡£
\end{eqnarray}
The functional relation obtained by this argument is consistent with the formulas obtained by Son and Spivak \cite{Son13prb} and Burkov \cite{Burkov14prl-chiral}. The conductivity increases with $B^2$, giving rise to a negative magnetoresistance.
Because the nontrivial Berry curvature diverges at the Weyl nodes, the positive conductivity will increase with decreasing Fermi wave vector and carrier density. In three dimensions, the carrier density $n$ is proportional to $k_F^3$, so
\begin{eqnarray}
\delta \sigma (B) \propto \frac{B^2}{n^{2/3}}.
\end{eqnarray}

Therefore, it is necessary to check three properties in order to verify a negative magnetoresistance from the nontrivial Berry curvature. (i) The angular dependence. Because of the $\mathbf{E}\cdot \mathbf{B}$ term in Eq. (\ref{Eq:Niu-eqn2}), the effect is maximized when the electric field is aligned with the magnetic field. Also, when the field is perpendicular to the current, the positive magnetoresistance from the Lorentz force can easily overwhelm the Berry-curvature negative magnetoresistance. (ii) The $B^2$ magnetic field dependence. (iii) The $n^{-2/3}$ carrier density dependence. So far, the first two properties have been verified by all the experiments in which the negative magnetoresistance is observed. In the experiment by Li et al. on a nanoribbon of Cd$_3$As$_2$ \cite{LiH16nc}, the carriers can be released by defects with increasing temperature, following an Arrhenius's law. The carrier density was extracted from two formulas. One is Kohler's rule $R(B_{\perp})=R_0[1+(\mu B_{\perp})^2]$, where $R(B_\perp)$ and $R_0$ are the resistance in the presence and absence of a perpendicular magnetic field $B_\perp$, and $\mu$ is the mobility. This can give a rough estimate of the mobility $\mu$, which is then put into the zero-field resistivity $\rho=1/n e\mu$ to yield the carrier density $n$ approximately. In a temperature window between 50K and 150K, the weak anti-localization does not play a role, and the change in the negative magnetoresistance can be assumed to be mainly from the change of the carrier density because it is a semiclassical conductivity contribution. The experiment shows that the coefficient in front of the negative magnetoresistance can be well fitted by $  B^2/n^{2/3}$.
In the experiment by Zhang et al. \cite{ZhangCL16nc}, the carrier density dependence was checked by comparing the results from different samples.

\section{Strong magnetic fields: the quantum limit}\label{Sec:QL}

\subsection{Argument of negative magnetoresistance in the quantum limit}

According to Nielsen and Ninomiya \cite{Nielsen83plb}, the original proposal for realizing the chiral anomaly in lattices is in the quantum limit of a 3D semimetal.
They started with a one-dimensional model in which two chiral energy bands have linear dispersions and opposite velocities.
An external electric field can accelerate electrons in one band to higher energy levels, in this way, charges are ``created". In contrast, in the other band, which has the opposite velocity, charges are annihilated. The chiral charge, defined as the difference between the charges in the two bands, therefore is not conserved in the electric field. This is literally the chiral anomaly. As one of the possible realizations of the one-dimensional chiral system, they then proposed to use the $\nu=0$ Landau bands of a three-dimensional semimetal, and expected ``the longitudinal magneto-conduction becomes extremely strong". In other words, the magnetoresistance of the 0th Landau bands in semimetals is the first physical quantity that was proposed as one of the signatures of the chiral anomaly.

In the quantum limit, only
the band of $\nu=0$ is partially filled. In this case the transport
properties of the system are dominantly determined by the highly degenerate
$\nu=0$ Landau bands [the red curve in Fig.~\ref{Fig:Landau}
(a)]. It is reasonable to regard them as a bundle of one-dimensional
chains. Combining the Landau degeneracy $N_{L}$, the $z$-direction
conductance is approximately given by
\begin{eqnarray}
\sigma_{zz}=N_{L}\sigma_{\text{1D}},
\end{eqnarray}
where $\sigma_{\text{1D}}$ is the conductance for each one-dimensional
Landau band.

If we ignore the scattering between the states in the degenerate Landau
bands, according to the transport theory, the ballistic conductance
of a one-dimensional chain in the clean limit is given by
\begin{eqnarray}
\sigma_{\text{1D}}=\frac{e^{2}}{h},
\end{eqnarray}
then the conductivity is found as
\begin{eqnarray}\label{NMC-ballistic}
\sigma_{zz}=\frac{e^{2}}{h}\frac{eB}{h},
\end{eqnarray}
which is is linear in magnetic field $B$, giving a positive magnetoconductivity.

In most measurements, the sample size is much larger than the mean
free path, then the scattering between the states in the Landau bands
is inevitable, and we have to consider the other limit, i.e., the
diffusive limit. Usually, the scattering is characterized by a momentum
relaxation time $\tau$. According to the Einstein relation, the conductivity
of each Landau band in the diffusive limit is
\begin{eqnarray}
\sigma_{\text{1D}}=e^{2}N_{\text{1D}}v_{F}^{2}\tau,
\end{eqnarray}
where $v_{F}$ the Fermi velocity and the density of states for each
1D Landau band is $N_{\text{1D}}=1/\pi\hbar v_{F}$, then
\begin{eqnarray}\label{NMC-diffusive}
\sigma_{zz}=\frac{e^{2}}{h}\frac{eBv_{F}\tau}{\pi\hbar}.
\end{eqnarray}
If $v_{F}$ and $\tau$ are constant, one readily concludes that the
magnetoconductivity is positive and linear in $B$.

Recently, several theoretical works have formulated the negative magnetoresistance or positive magnetoconductivity in the quantum limit as one of the signatures of the chiral anomaly \cite{Son13prb,Gorbar14prb}, much similar to those in Eqs. (\ref{NMC-ballistic}) and (\ref{NMC-diffusive}).
In both cases, the positive magnetoconductivity arises because the
Landau degeneracy increases linearly with $B$. However, in the following, we will show that if $v_{F}$ and $\tau$
also depend on the magnetic field, the conclusion has to be reexamined.

\subsection{Disorder scattering}

One of the
convenient choices is the random Gaussian potential
\begin{eqnarray}\label{Eq:U-Gaussian}
U(\mathbf{r}) & = & \sum_{i}\frac{u_{i}}{(d\sqrt{2\pi})^{3}}e^{-|\mathbf{r}-\mathbf{R}_{i}|^{2}/2d^{2}},\label{U-Gaussian}
\end{eqnarray}
where $u_{i}$ measures the scattering strength of a randomly distributed impurity at $\mathbf{R}_{i}$, and $d$ is a parameter that determines the range of the scattering potential.
The Gaussian potential allows us to study the effect of the potential
range in a controllable way, which we find it crucial in the present
study. Now we have two characteristic lengths, the potential range
$d$ and the magnetic length $\ell_{B}$, which define two regimes,
the long-range potential regime $d\gg\ell_{B}$ and the short-range
potential limit $d\ll\ell_{B}$. Note that, for a given $d$ in realistic
materials, varying the magnetic field alone can cross between the
two regimes. Empirically, the magnetic length $\ell_{B}$ = 25.6 nm
/$\sqrt{|B|}$ with $B$ in Tesla. In the strong-field limit, e.g.,
$B>10$ T, the magnetic length $\ell_{B}$ becomes less than 10 nm,
it is reasonable to regard smooth fluctuations in materials as long-range.

For the scattering among the states on the Fermi surface of the $0$th Landau bands, the transport time can be found as
\begin{eqnarray}
\frac{\hbar}{\tau^{0,\text{tr}}_{k_F}}
&=& 2\pi \sum_{k_x',k_z'}
\langle |U^{0,0}_{k_x,k_F;k_x',k_z'}|^2 \rangle \delta(E_F-E^0_{k_z'}) \nonumber\\ &&\times (1-\frac{v^z_{0,k_z'}}{v_F}),
\end{eqnarray}
where $U^{0,0}_{k_x,k_F;k_x',k_z'}$ represents the scattering matrix elements calculated from Eq. (\ref{Eq:U-Gaussian}) and $\langle ... \rangle$ means the impurity average \cite{ZhangSB16njp}.

\subsection{Negative magnetoconductivity with Delta potential}

The delta potential means $d\rightarrow 0$ in Eq. (\ref{Eq:U-Gaussian}). In this case, the transport time is the same as the scattering time \cite{Lu15Weyl-shortrange}.
By considering the magnetic field dependence of the scattering time, we find that
in the strong-field limit ($B\rightarrow\infty$),
\begin{eqnarray}\label{tau-00-strongB}
\tau  & = & \frac{\hbar^{2}v_{F}^{0}\pi\ell_{B}^{2}}{V_{\text{imp}}}.
\end{eqnarray}
Here we suppress the correction $\Lambda$ , because it cancels in $\sigma_{zz}$ \cite{Lu15Weyl-shortrange}.
The scattering time can be put into Eq. (\ref{NMC-diffusive}) to give the conductivity in the strong-field limit as
\begin{eqnarray}
\sigma_{zz,0}^{sc} & = & \frac{e^{2}}{h}\frac{(\hbar v_{F}^{0})^{2}}{V_{\text{imp}}}.\label{Eq:sigma-zz-0-v}
\end{eqnarray}
Notice that the Landau degeneracy in the scattering time cancels with
that in Eq. (\ref{NMC-diffusive}), thus the magnetic field dependence
of $\sigma_{zz,0}^{sc}$ is given by the Fermi velocity $v_{F}^{0}$.
When ignoring the magnetic
field dependence of the Fermi velocity, a $B$-independent conductivity
was concluded, which is consistent with the previous work in which the velocity is constant \cite{Aji12prbrc}.
We find the magnetic field dependence of the Fermi velocity can lead to different scenarios of positive and negative magnetoconductivity.

(i) Weyl semimetal with fixed carrier density. In a strong field the Fermi velocity or the Fermi energy is given
by the density of charge carriers and the magnetic field \cite{Abrikosov98prb}.
We assume that an ideal Weyl semimetal is the case that the Fermi
energy crosses the Weyl nodes, all negative bands are fully filled
and the positive bands are empty. In this case $\hbar v_{F}^{0}=2M_{1}k_w$.
An extra doping of charge carriers will cause a change of electron
density $n_{0}(>0)$ in the electron-doped case or hole density $n_{0}(<0)$
in the hole-doped case. The relation between the Fermi wave vector
and the density of charge carriers is given by
\begin{equation}
n_{0}=2 N_{L}\times\frac{k_{F}^{0}-k_w}{2\pi}
\end{equation}
This means that  the  Fermi wave vector is determined by the density
of charge carriers $n_{0}$ and magnetic field $B$,
\begin{eqnarray}\label{kF-weyl-fixn0}
k_{F}^{0} & = & k_w+\pi n_{0}h/eB
\end{eqnarray}
or $k_{F}^{0}=k_w+2\pi^{2}n_{0}\ell_{B}^{2}$. Thus the Fermi velocity
is also a function of $B$, $\hbar v_{F}^{0}=2M_{1}k_{F}^{0},$ and
\begin{eqnarray}
\sigma_{zz,0}^{sc}=\sigma_{N}\left[1+\mathrm{sgn}(n_{0})\frac{B_{c}}{B}\right]^{2}.
\label{Eq:sigma-zz-0-B2}
\end{eqnarray}
where the characteristic field $B_{c}=\pi\left|n_{0}\right|h/ek_w$.
A typical order of $B_{c}$ is about 10 Tesla for $n_{0}$ of 10$^{17}$/cm$^{3}$. $\sigma_{zz,0}^{sc}$ is
constant for the undoped case of $n_{0}=0$, and
\begin{eqnarray}
\sigma_{N}=\frac{e^{2}}{h}\frac{4M_{1}^{2}k_w^{2}}{V_{\text{imp}}}\label{sigma-N}
\end{eqnarray}
is the conductivity of the undoped case, and is independent of magnetic
field. Thus the magnetoconductivity is always negative in the electron-doped
case while always positive in the hole-doped regime.

(ii) Weyl semimetal with fixed Fermi energy.
In the case that the Fermi energy is fixed, $(\hbar v_{F}^{0})^{2}=4M_{1}(E_{F}-eM_{1}B/\hbar+M_{0})$,
and we have
\begin{eqnarray}
\sigma_{zz,0}^{sc} & = & \frac{e^{2}}{h}\frac{4M_{1}(E_{F}-eM_{1}B/\hbar+M_{0})}{V_{\text{imp}}},\label{Eq:MC-EF}
\end{eqnarray}
then the magnetoconductivity is always negative and linear in $B$.

(iii) Paramagnetic semimetal. For the Dirac semimetal or paramagnetic semimetal described by Eq. (\ref{Eq:H-Dirac}), there are two branches of $\nu=0$
bands, with the energy dispersions $E_{k_{z}}^{0\uparrow}=\omega/2+\Delta_{p}-M_{0}+M_{1}k_{z}^{2}$
and $E_{k_{z}}^{0\downarrow}=-\omega/2-\Delta_{s}+M_{0}-M_{1}k_{z}^{2}$
for $H(\mathbf{k})$ and $H^{*}(-\mathbf{k})$, respectively. In the absence of inter-block velocity, the longitudinal
conductance along the $z$ direction is approximately a summation
of those for two independent Weyl semimetals.
First, we consider the Fermi energy cross both bands $0\uparrow$
and $0\downarrow$. Using Eq. (\ref{Eq:MC-EF}), the $z$-direction
conductivity is found as
\begin{eqnarray}
\sigma_{zz,0}^{sc} & = & \sigma_{zz,0\uparrow}^{sc}+\sigma_{zz,0\downarrow}^{sc}\nonumber \\
 & = & \frac{e^{2}}{h}\frac{8M_{1}}{V_{\text{imp}}}[M_{0}-\frac{eM_{1}B}{\hbar}-\frac{\mu_{B}(g_{p}+g_{s})B}{4}],
\end{eqnarray}
or using $\sigma_{N}$ defined in Eq. (\ref{sigma-N}),
\begin{eqnarray}
\sigma_{zz,0}^{sc} & = & 2\sigma_{N}[1-\frac{eB}{\hbar k_w^{2}}-\frac{\mu_{B}(g_{p}+g_{s})B}{4M_{0}}].\label{Eq:MC-Dirac}
\end{eqnarray}
In this case we have a negative linear $B$ magnetoconductivity, when
the Fermi energy crosses both $E_{k_{z}}^{0\uparrow}$ and $E_{k_{z}}^{0\downarrow}$.
With increasing magnetic field, the $0\uparrow$ bands will shift
upwards and the $0\downarrow$ bands will shift downwards. Beyond
a critical field, the Fermi energy will fall into either $0\uparrow$
or $0\downarrow$ bands, depending on whether the carriers are electron-type
or hole-type. If the carrier density is fixed, the Fermi wave vector
in this case does not depend on $k_w$ as that in Eq. (\ref{kF-weyl-fixn0}),
but
\begin{eqnarray}
k_{F}^{0} & = & \frac{\pi n_{0}h}{eB}
\end{eqnarray}
or $k_{F}^{0}=2\pi^{2}n_{0}\ell_{B}^{2}$. In this case, with increasing
magnetic field, the Fermi energy will approach the band edge and the
Fermi velocity always decreases. Using Eq. (\ref{Eq:sigma-zz-0-v}),
\begin{eqnarray}
\sigma_{zz,0}^{sc} & = & \frac{e^{2}}{h}\frac{4\pi^{2}h^{2}M_{1}^{2}n_{0}^{2}}{V_{\text{imp}}e^{2}B^{2}},\label{sigma-zz-dirac-largeB}
\end{eqnarray}
which also gives negative magnetoconductivity that is independent
on the type of carriers. Note that in the Weyl semimetal TaAs with broken inversion symmetry,
where the Weyl nodes always come in even pairs because of time-reversal
symmetry \cite{Weng15prx,Huang15nc,Lv15prx,Xu15sci-TaAs}, the situation
is more similar to that for the Dirac semimetal and the magnetoconductivity
does not depend on the type of carriers and may be described by a
generalized version of Eqs. (\ref{Eq:MC-Dirac}) and (\ref{sigma-zz-dirac-largeB}).

\subsection{Positive linear magnetoconductivity and zero-field minimum conductivity at half filling of a Weyl semimetal}

\begin{figure}
\centering \includegraphics[width=0.45\textwidth]{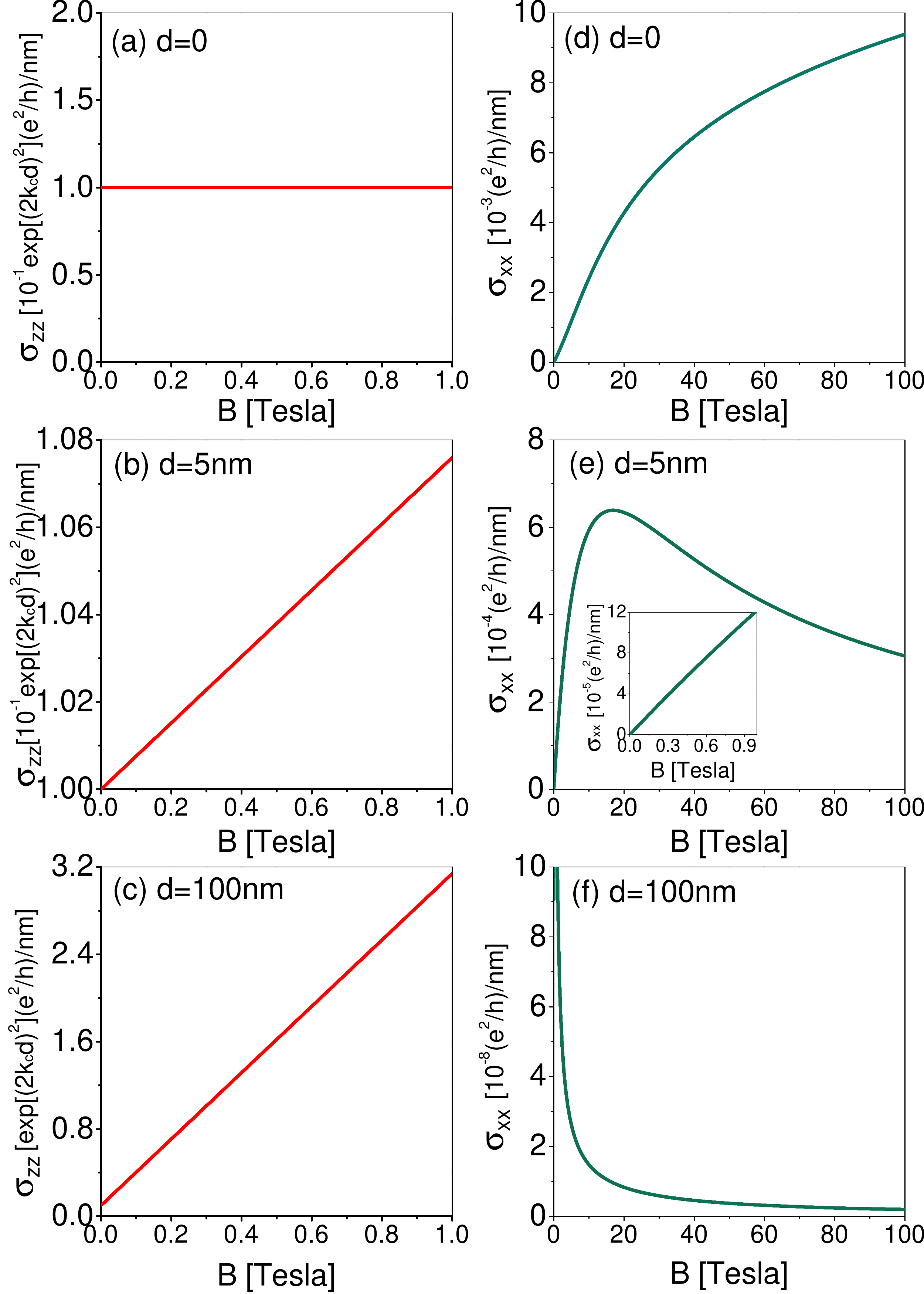}
\protect\caption{The longitudinal conductivity $\sigma_{zz}$ and transverse conductivity
$\sigma_{xx}$ of the Weyl semimetal in the $\hat{\bf{z}}$-direction magnetic
field $B$ for different potential ranges. The shared parameters: $k_w=0.1$/nm,
$M=5$ eV$\cdot$nm$^{2}$, $A=1$ eV$\cdot$nm, $V_{\text{imp}}=10$ (eV)$^{2}$$\cdot$nm$^{3}$. Adapted from Ref. \cite{ZhangSB16njp}.}
\label{Fig:Weyl-Gaussian}
\end{figure}

With the random Gaussian potential, we can find the transport time as well as the conductivity. In particular, at the Weyl nodes the transport time is obtained as \cite{ZhangSB16njp}
\begin{eqnarray}\label{tau-0-gaussian-twonode-B}
\frac{\hbar}{\tau^{0,\text{tr}}_{k_w}}
&=& \frac{V_{\text{imp}}}{2\pi Mk_w}
\frac{ e^{- 4d^2 k_w^2}}{2d^2+\ell_B^2},
\end{eqnarray}
and hence the longitudinal conductivity
\begin{eqnarray}
\sigma_{zz}(B) & = & \frac{e^{2}}{h}\frac{(2Mk_w)^{2}( 2d^{2}+\ell_{B}^{2})}{V_{\text{imp}}\ell_{B}^{2} }e^{4 d^{2}k_w^{2}}, \label{sigma-zz-gaussian-vF}
\end{eqnarray}
where $V_{\text{imp}}\equiv \sum_i u_i^2/V $ measures the strength of the scattering and $V= L_xL_yL_z$ is the volume of the system. $L_{x,y,z}$ are the sizes of the system along the $\hat{\mathbf{x}}$, $\hat{\mathbf{y}}$ and $\hat{\mathbf{z}}$ directions, respectively.  This conductivity is generated by the inter-node scattering with a momentum
transfer of $2k_w$. As the magnetic field goes to zero, the magnetic length diverges and  $d/\ell_B\rightarrow 0$, and Eq. (\ref{sigma-zz-gaussian-vF}) gives a minimum conductivity
\begin{eqnarray}
\sigma_{zz}(0) & = & \frac{e^{2}}{h}\frac{ 4(Mk_w)^{2}}{V_{\text{imp}}  }e^{4d^{2}k_w^{2}},\label{sigma-zz-gaussian-minimum}
\end{eqnarray}
even though the DOS vanishes  at the Weyl nodes at zero
magnetic field. A similar result was found in the absence of the Landau
levels \cite{Ominato14prb}.

According to $d$, we have two cases. (1) In the short-range limit,
$d=0$, then $\sigma_{zz}$ does not depend on the magnetic field,
giving a zero magnetoconductivity, which recovers the result for the
delta potential \cite{Lu15Weyl-shortrange,Goswami15prb}. (2) As long
as the potential range is finite, i.e., $d>0$, we can have a magnetoconductivity.
Using Eq.~(\ref{sigma-zz-gaussian-vF}),
\begin{eqnarray}
\Delta\sigma_{zz}(B) & \equiv & \dfrac{\sigma_{zz}(B)-\sigma_{zz}(0)}{\sigma_{zz}(0)}=\dfrac{B}{B_{0}},\label{sigma-zz-gaussian-B}
\end{eqnarray}
where $B_{0}=\hbar/2ed^{2}$. Thus the magnetoconductivity is given
by the range of impurity potential, and independent of the
model parameters. This means that we have a positive linear $\hat{\mathbf{z}}$-direction
magnetoconductivity for the Weyl semimetal. A finite
carrier density $n_{0}$ can drive the system away from the Weyl nodes,
then $k_w$ in Eq.~(\ref{sigma-zz-gaussian-vF}) is to be replaced
by $k_{F}=k_w+\mathrm{sgn}(M)2\pi^{2}\ell_{B}^{2}n_{0}$. The finite
$n_{0}$ can vary the linear-$B$ dependence, but a strong magnetic
field can always squeeze the Fermi energy to $k_w$, and recover
the linear magnetoconductivity.

A linear-$B$ magnetoconductivity arising from the Landau degeneracy has been obtained before \cite{Son13prb,Gorbar14prb}, based on the assumption that the transport time and Fermi velocity are constant. However, in the present case, we have taken into account the magnetic field dependence of the transport time, and thus the $B$-linear magnetoconductivity here has a different mechanism as a result of the interplay of the Landau degeneracy and impurity scattering. Also, in the presence of the charged impurities, a $B^{2}$ magnetoconductivity can be found in the quantum limit \cite{Goswami15prb}. A $B^2$ magnetoconductivity can also be found in the semiclassical
limit \cite{Son13prb,Burkov14prl-chiral}. Numerically, a positive magnetoconductivity is also found for the long-range disorder, although the system
tends to have negative magnetoconductivity for the weak short-range disorder \cite{ChenCZ16prb}.

\subsection{Transverse magnetoconductivity \label{Sec:Sxx}}
 When electric and magnetic
fields are perpendicular to each other, the changing rate of density
of charge carriers near each node vanishes. In this case, because
the Landau bands in the $\hat{\mathbf{z}}$-direction magnetic field only disperse
with $k_{z}$, the effective velocity along the $\hat{\bf{x}}$ direction $v_{x}=\partial E_0/\hbar \partial k_{x}=0$.
The leading-order $\hat{\bf{x}}$-direction conductivity arises from the inter-band
velocity and the scatterings between the 0th bands with the bands of $1\pm$,
which are higher-order perturbation processes. Thus the transverse conductivity is usually much smaller than the longitudinal conductivity.

There are three cases as shown in Figs.~\ref{Fig:Weyl-Gaussian} (d)-(f). At
$d=0$, $\sigma_{xx}$ reduces to the result for the delta potential
and $\sigma_{xx}\propto B$, a linear magnetoconductivity as $\sigma_{zz}$,
but much smaller \cite{Lu15Weyl-shortrange}. In the long-range potential
limit $d\gg\ell_{B}$, we have $\sigma_{xx}\sim1/B$, which gives
a negative magnetoconductivity. For a finite potential range $d$,
we would have a crossover of $\sigma_{xx}$ from $B$-linear to $1/B$
dependence. Alternatively, as shown in Fig.~\ref{Fig:Weyl-Gaussian} (e), for
a finite $d$ ($=5$ nm) comparable to the magnetic length $\ell_{B}$,
we have a crossover of $\sigma_{xx}$ from a linear-$B$ dependence
in weak fields to a $1/B$ dependence in strong fields. While at $d=0$
and $d\gg \ell_B$, we have the two limits as shown in Figs. \ref{Fig:Weyl-Gaussian} (d) and (f), respectively. For shorter $d$, a larger critical magnetic field for the crossover is needed. Figure \ref{Fig:Weyl-Gaussian}
also shows that the conductivity is larger for shorter $d$, so the $1/B$ transverse magnetoconductivity in the long-range limit may not survive when there are additional short-range scatters.

In particular, in Fig.~\ref{Fig:Weyl-Gaussian} (f), $\sigma_{xx}\propto1/B$
in the long-range potential limit. In the field perpendicular to the
$x\text{-}y$ plane, there is also a Hall conductivity $\sigma_{yx}=\mathrm{sgn}(M)(k_w/\pi)e^{2}/h+en_{0}/B$,
where the first term is the anomalous Hall conductivity and the second
term is the classical conductivity. In weak fields, the classical Hall
effect dominates, then both $\sigma_{xx}$ and $\sigma_{yx}$ are
proportional to $1/B$, and the resistivity $\rho_{xx}=\sigma_{xx}/(\sigma_{xx}^{2}+\sigma_{yx}^{2})$
is found to be linear in $B$. Note that here the linear MR in perpendicular fields has a different scenario compared to the previous works \cite{Abrikosov98prb,Song15prb}. Abrikosov used the Hamiltonian $v \vec{k}\cdot\vec{\sigma}$
with linear dispersion and modelled the disorder by the screened Coulomb potential under the random phase approximation \cite{Abrikosov98prb}. Song \emph{et al}. discussed a semiclassical mechanism \cite{Song15prb}.

\section{Remarks and Perspective}\label{Sec:remark}

In summary, we have systematically studied the quantum transport in topological semimetals, including the weak (anti-)localization, negative magnetoresistance, and the magneto-transport in the quantum limit.

A single valley of Weyl fermions has the weak anti-localization, while a single valley of double-Weyl fermions has the weak localization. In the presence of strong intervalley effects, both Weyl and double-Weyl semimetals have the weak localization. The interplay of electron-electron interaction and disorder scattering can also give rise to a tendency to localization for Weyl fermions. For Weyl and double-Weyl semimetals, we derived a magnetoconductivity formula, which connects the $B^2$ behavior near zero field and $\sqrt{B}$ behavior in stronger fields, for the weak (anti-)localization in three dimensions. Our formula of magnetoconductivity can be used for a systematic study of the transport experiments on topological semimetals.

We review the experiments on the negative magnetoresistance in topological semimetals.
Using the semiclassical equation that includes the anomalous velocity induced by the Berry curvature, we show the relation between the magnetic monopole and the negative magnetoresistance. The negative magnetoresistance is shown to diverge according to $1/n^{2/3}$, where $n$ is the carrier density. Therefore, demonstrating the carrier density dependence of the negative magnetoresistance is a crucial step to show the nontrivial topological properties of topological semimetals.

In the quantum limit, we show that the negative magnetoresistance is not a compelling signature of the chiral anomaly. The sign of the magnetoresistance in the quantum limit depends the details of the disorder and band dispersions. We give the conditions of the negative magnetoresistance. For long-range Gaussian potential and at half filling, we can have a linear magnetoconductivity. We also find a minimal conductivity at the Weyl nodes, although the density of states vanishes.

Finally, we remark on the possible future works.
The weak (anti-)localization theories for nodal-line and drumhead semimetal could be interesting topics. It is known that
the ``chiral anomaly" could give a positive magnetoconductivity \cite{Son13prb,Kim14prb,Burkov14prl-chiral,Gorbar14prb}. A double-Weyl semimetal is also expected to have a negative magnetoresistance.
So far, most theories in the quantum limit employ the Born approximation, e.g., the quantum linear magnetoresistance \cite{Abrikosov98prb}. When the magnetic length becomes much shorter than the range of the disorder potential, electrons may be scattered by the same impurity for multiple times. The Born approximation contains the correlation of two scattering events by the same impurity \cite{Mahan1990}. In this situation, the validity of the Born approximation was questioned in two dimensions \cite{Raikh93prb,Murzin00pu}.
In three dimensions, it is still unclear whether the correlation of two scattering events in the Born approximation is the building block for the multiple scattering under extremely strong magnetic fields \cite{Song15prb,Pesin15prb}. The treatment beyond the Born approximation will be a challenging topic for three-dimensional systems under extremely strong magnetic fields. Recently, a linear and unsaturated magnetoresistance has been observed in many topological semimetals  \cite{Liang15nmat,Feng15prbrc,He14prl,Zhao15prx,Cao15nc,Shekhar15np,Narayanan15prl,Li16np,ZhangCL16nc,HuangXC15prx,Xiong15sci,YangXJ15arXiv,ZhangC15arXiv,LiCZ15nc,LiH16nc}. The origin of the linear magnetoresistance remains elusive and has attracted many theoretical works in the classical regime \cite{Parish03nat,Alekseev15prl,Ramakrishnan15prb,Pan15arXiv,Song15prb} and in the quantum regime \cite{Abrikosov98prb}. The theory of the linear magnetoresistance will still be an interesting topic. The superconductivity has been observed around the point contact region on the surface of Cd$_3$As$_2$ crystals \cite{WangH16nmat,Aggarwal16nmat}, which may inspire more explorations.

\acknowledgments
We thank fruitful discussions with Xin Dai, Hongtao He, Shuang Jia, Hui Li, Titus Neupert, Chunming Wang, Jiannong Wang, Suyang Xu, Hong Yao, Chenglong Zhang, Songbo Zhang.
This work was supported by the National Key R \& D Program (Grant No. 2016YFA0301700), the National Science Foundation of China (Grant No. 11574127) and the Research Grant Council, University Grants Committee, Hong Kong (Grant No. 17301116), and the National Thousand-Young-Talents Program of China.

%\bibliography{refs-transport}

%

%\end{CJK}
\end{document}